\documentclass[10pt,letterpaper]{article}
\usepackage[top=0.85in,left=0.75in,footskip=0.75in,marginparwidth=2in]{geometry}

\usepackage[utf8]{inputenc}

\usepackage{cite}

\usepackage{nameref,hyperref}

\usepackage[right]{lineno}

\usepackage{microtype}
\DisableLigatures[f]{encoding = *, family = * }

\usepackage{changepage}

\usepackage[aboveskip=1pt,labelfont=bf,labelsep=period,singlelinecheck=off]{caption}

\makeatletter
\renewcommand{\@biblabel}[1]{\quad#1.}
\makeatother

\usepackage{lastpage,fancyhdr,graphicx}
\usepackage{epstopdf}
\fancyheadoffset[L]{2.25in}
\fancyfootoffset[L]{2.25in}

\usepackage{color}

\definecolor{Gray}{gray}{.25}

\usepackage{graphicx}

\usepackage{sidecap}

\usepackage{wrapfig}
\usepackage[pscoord]{eso-pic}
\usepackage[fulladjust]{marginnote}
\reversemarginpar

\usepackage{amsmath}

\usepackage[scanall]{psfrag}

\begin{document}
\vspace*{0.35in}

% title goes here:
\begin{flushleft}
{\Large
\textbf\newline{Properties of Kinetic Transition Networks for Atomic Clusters and Glassy Solids}
}
\newline
% authors go here:
\\
John W.\ R.\ Morgan\textsuperscript{1},
Dhagash Mehta\textsuperscript{2},
David J.\ Wales\textsuperscript{3,*}
\\
\bigskip
\bf{1} Department of Chemical Engineering, University of Michigan, Ann Arbor, MI 48109-2136, USA. E-mail: jwrm@umich.edu
\\
\bf{2} Department of Applied and Computational Mathematics and Statistics, University of Notre Dame, Notre Dame, IN 46556, USA. E-mail: dmehta@nd.edu
\\
\bf{3} University Chemical Laboratories, Lensfield Road, Cambridge CB2 1EW, UK. E-mail: dw34@cam.ac.uk
\\
\bigskip
* Corresponding author

\end{flushleft}

\section*{Abstract}
A database of minima and transition states corresponds to a network
where the minima represent nodes and the transition states correspond to edges
between the pairs of minima they connect via steepest-descent paths. 
Here we construct networks 
for small clusters bound by the Morse potential for a selection of physically
relevant parameters, in two and three dimensions. The properties of these
unweighted and undirected networks are analysed to examine two
features: whether they are small-world, where the shortest path between
nodes involves only a small number or edges; and whether they are
scale-free, having a degree distribution that follows a power law. 
Small-world character is present, but statistical tests show that a
power law is not a good fit, so the networks are not scale-free. 
These results for clusters are compared with the corresponding properties for 
the molecular and atomic structural glass formers 
ortho-terphenyl and binary
Lennard-Jones. These glassy systems do not show small-world properties,
suggesting that such behaviour is linked to the structure-seeking landscapes of the
Morse clusters.

% now start line numbers
%\linenumbers

\section{Introduction}
\label{sec:intro}

The potential energy surface (PES)\cite{EnergyLandscapes} of an
atomic cluster corresponds to the energy as a function of the coordinates specifying
the configuration. The most interesting points on the surface are usually
the local minima and transition states, which are stationary points
where the gradient of the potential is zero.
For local minima, the potential energy rises for any infinitesimal displacement of
internal coordinates, while
for transition states
there is a unique negative Hessian (second derivative matrix)
eigenvalue.\cite{MurrellL68} Treating the PES as a network can provide
insight into the overall structure of the energy landscape. 
Here the network in question is formed by
considering minima as the nodes and transition states as edges between the
minima they connect via steepest-descent paths.\cite{StillingerW82,StillingerW84} 
Two key questions are then
whether the network is small-world and scale-free.

The degree of a node is the number of edges connected to it. The degree
distribution is then defined as the number of nodes with given
degrees.\cite{Networks} A path is a sequence of nodes connected by edges, with
the length of the path being the number of edges it contains. The shortest path
length gives the minimum number of edges between a pair of nodes, and the
average shortest path length is the shortest path length averaged over all
pairs of nodes. The clustering coefficient is the ratio of the number of
connections between neighbours of a node to the number of such connections
that could exist.\cite{WattsS98}

Two useful references are provided by models based on a lattice graph and a random graph. 
Random graphs have a
small clustering coefficient and a slowly growing average shortest path
length.\cite{Networks, FronczakFH02}
In contrast, lattice graphs, which are composed of a
regular array of nodes with edges only between nearest-neighbours in space,
have a comparatively large average shortest path length and a larger clustering
coefficient.\cite{WattsS98} Small-world networks were introduced by Watts and
Strogatz\cite{WattsS98} and defined as networks that show a high degree of
local clustering behaviour, similar to a lattice graph, but also a short path
length, even between distant nodes, as exhibited by random graphs. The name comes
from experiments performed by Milgram with letter passing, which demonstrated that
most citizens of the USA are separated from each other by a surprisingly small
number of social contacts.\cite{Milgram67} This result is now known in terms of
``six degrees of separation'' from the estimated
average path length.\cite{SixDegrees} Watts and Strogatz showed that networks
from a wide variety of areas show small-world behaviour, including the neural
network of \textit{C. elegans}, the power grid of the Western USA, and the
collaborations of film actors (linked to the concept of the Bacon
number)\cite{BaconNumber}.

A scale-free network has a degree distribution with a
power law tail, as defined by Barab\'{a}si and Albert.\cite{BarabasiA99, AlbertB02}. This
distribution implies there are a small number of nodes with a very high degree,
called hubs, and more nodes with a smaller degree, acting as local hubs, down
to nodes with only a few connections, in a hierarchical fashion. 
Power laws were fitted for a variety of available networks, including the World Wide Web and
the film actor collaboration graph, and a preferential attachment
model was suggested to explain how the behaviour arises. Following this work, power laws
were fitted to many other networks, such as the interactions between solar
magnetic loops leading to a solar flare,\cite{HughesPDHM03} and aftershocks of
earthquakes\cite{BaiesiP04}. However, few of these studies used robust
statistical methods to determine whether or not a power law was a good fit to
the data, and many of the results have been called into question by Clauset
\textit{et al.}\cite{ClausetSN09}

Doye and Massen\cite{Doye02, DoyeM05} studied Lennard-Jones (LJ) clusters and
concluded that the networks were both small-world and scale-free. Small-world
behaviour is believed to have implications for self-assembly, as it suggests
that the global minimum can be reached from anywhere on the PES by a relatively
short transition pathway. One would therefore expect that systems with a
funnel-like PES, characteristic of structure-seekers, may display small-world
behaviour.\cite{CarrW08} Scale-free behaviour was more surprising, as there is no obvious
analogy to the addition of new nodes in the preferential attachment model, the
most straightforward route to a power law distribution. However, noticing a
correlation between minima with a high degree and a low potential energy,
it was suggested that due to the larger basins of attraction for low energy
minima, any other minimum was more likely to be connected to a low energy
minimum than a high energy one, thus establishing an analogy to preferential
attachment. This possibility was further investigated using Apollonian
packings\cite{DoyeM05b}, concluding that the contacts between discs in the 2D
Apollonian packing form a scale-free network with a spatial distribution that
may resemble the catchment basins of a PES.

Other energy landscapes have been studied with a view to analysing the same
properties. Protein folding networks have been considered by Rao and
Caflisch\cite{RaoC04}. They used molecular dynamics to generate snapshots of
the structure and then derived a conformation based on the secondary structure
each residue belonged to. These snapshots do not precisely correspond to minima
on the PES, so the formation of the network is not equivalent to the LJ results
considered above, but the idea is similar. A power law was fitted to the degree
distribution with the conclusion was that the network was scale-free, drawing
attention to the hubs present in the network. 

Bowman and Pande
studied a Markov state model for another protein folding landscape.\cite{BowmanP10} They
noted that native states were hubs and suggested that this was
characteristic of a scale-free network, but without consideration of the degree
distribution. Similarly, Chakraborty \textit{et al.} considered an RNA folding
network, noting that it may be scale-free due to the presence of hubs, but without
providing the degree distribution.\cite{ChakrabortyCGW14} It is, however,
possible to construct networks with hubs that are not scale-free, the simplest
example being the star network,\cite{Networks} in which there is one central
node to which all others are attached and there no other edges. Therefore,
although scale-free networks do contain hubs, the presence of hubs is not in
itself enough evidence to conclude that a network is scale-free. Recently, Mehta
\textit{et al.}\cite{MehtaCCKW16} constructed stationary point databases
for the Thomson problem\cite{Thomson04} up to 150 particles. They
demonstrated that the networks show small-world behaviour, and that the
presence of hubs suggested the possibility of the networks being scale-free,
but acknowledged that further statistical testing was required.
Recently, an analysis of networks of minima for machine learning problems has also
been initiated.\cite{BallardDMMSSW17}

Potential energy surfaces for glassy systems have a large number of competing 
low-lying amorphous minima
with similar energies,
separated by high barriers.\cite{EnergyLandscapes}. Here we consider two examples: bulk ortho-terphenyl
(OTP), which is often represented by a course-grained model consisting of a LJ
site at the centre of each benzene ring;\cite{NiblettSSW16} and bulk binary
Lennard-Jones (BLJ), in which there are two different types of particles,
interacting via LJ potentials with different
parameters.\cite{deSouzaW06,deSouzaW08} If the small-world
behaviour observed for atomic clusters is the result of a funnelled landscape, then
we anticipate that glassy systems will be different.

The Morse potential \cite{Morse29} is a widely used pairwise representation
for modelling atomic clusters.\cite{BraierBW90,DoyeWB95,MainzB96,DoyeW96d,DoyeW96a,DoyeW97a,ChengY07b,FengCL09}
The functional form is \begin{equation}
\label{eq:morse}
V_{M}\left(R\right) = \epsilon e^{\rho(1-R/R_{e})}(e^{\rho(1-R/R_{e})}-2),
\end{equation}
where the potential between two particles, in terms of the distance
between them, $R$, is parameterised by the pair equilibrium distance, $R_{e}$, 
by $\epsilon$, the pair well depth at $R = R_{e}$, and by the range
parameter $\rho$. Taking $\epsilon = 1$ and $R_{e} = 1$ sets the energy and
length units. The range relative to the particle diameter can be adjusted to
represent different systems: $\rho = 3$ is a long range, appropriate for sodium
atoms,\cite{GirifalcoW59} while $\rho = 14$ is a short range, suitable for
$\mathrm{C}_{60}$;\cite{DoyeW96a} $\rho = 30$ is a very short range, which has
been used for modelling colloids.\cite{Wales10b, MalinsWETR09, TaffsMWR10,
CalvoDW12, MorganW14} 

The Morse potential is pairwise additive so the total energy of the system is a sum over the
interaction between all pairs of particles. This approximation neglects
many-body interactions, which can be important in some systems, especially
when the particle interaction range is
large.\cite{MerrillSK09} The pair approximation has been successfully used
to predict experimental properties, in colloids for example, when the range is small. \cite{Wales10b}

In the present work we explore 
the PES for small atomic clusters bound by the Morse potential with a range of values for $\rho$,
namely $3,\ 6,\ 10,\ 14$ and $30$. Our databases are almost complete, 
meaning that they were expanded until no new minima were found, 
but the search methods used do not rigorously guarantee that all the minima have been located. 
Disconnectivity graphs are used to visualise the potential energy
landscape.\cite{BeckerK97, WalesMW98} Previously, global minima and some low-lying minima have 
been described, \cite{DoyeWB95, DoyeW96d, DoyeW96a, MainzB96,
DoyeW97a, ChengY07b, ChengY07c, FengCL09, CalvoDW12} but attempts to generate
relatively complete databases have been made only for a few specific cases.
\cite{BraierBW90, MorganW14}

\section{Methods}
\label{sec:methods}
\subsection{Generating Databases}
Clusters in two and three dimensions bound by the Morse potential were
analysed. For each number of particles $N$ and range $\rho$ a database of
potential energy minima and the transition states 
connecting them was generated. First, the global minimum and other low-lying
minima were located by basin-hopping global optimisation\cite{EnergyLandscapes,
LiS87, LiS88, WalesD97, WalesS99, WalesB06} as implemented in
\texttt{\textsc{GMIN}}.\cite{GMIN} $10^{4}$ basin-hopping steps were run from each of three
random starting configurations. The twenty lowest minima from each run,
or all minima if there were fewer than twenty, were then connected \textit{via}
transition states located using the doubly-nudged\cite{TrygubenkoW04,
TrygubenkoW04a, SheppardTH08} elastic band\cite{MillsJ94, MillsJS95, NEB,
HenkelmanUJ00, HenkelmanJ00} and eigenvector-following\cite{MunroW99}
algorithms in \texttt{\textsc{OPTIM}}.\cite{OPTIM} These minima and transition
states were imported into \texttt{\textsc{PATHSAMPLE}},\cite{PATHSAMPLE} at
which point duplicate minima and transition states were removed. Additional
single-ended transition state searches\cite{WalesDMMW00, MunroW99, KumedaMW01}
and searches using the UNTRAP\cite{StrodelWW07} procedure were performed to
expand the database. After no more transition states were located, a
further 50 single-ended searches were performed for each minimum. At this point
we can be reasonably confident that practically all the minima and transition states have
been found.

For each database, a network was constructed by considering the minima as nodes
with edges between any two minima directly connected by a transition state. The
barrier height for the transition state was ignored and the path through the
transition state can be followed in either direction, so the network was
treated as undirected and unweighted. The network was analysed using the Python
NetworkX package.\cite{NetworkX} A further check on the completeness of the
database was to ensure that all minima were connected to each other by one or
more transition states, \textit{i.e.}\ that the network was connected. Finally,
transition states connecting the same pair of minima as an earlier pathway were
removed, to leave at most one transition state connecting any pair. Transition
states that connected permutation-inversion isomers of the same minimum, which
correspond to degenerate rearrangements,\cite{LeoneS70} were also removed.

We chose to study clusters with $\rho = 3, 6, 10, 14$ and $30$, in both two and
three dimensions. These range parameters correspond to previous values used for
the Morse potential and give a good spread of physically relevant pair
potentials. The number of particles was chosen to produce databases with
between several hundred and a few thousand minima. Selecting these sizes meant
we could be confident of generating essentially complete databases in a 
reasonable time, improving the reliability of statistical tests. The symbol
M$^{xD}_{N}$ is used to refer to a Morse cluster in $x$ dimensions with $N$
particles.

\subsection{Degree Distribution}
A node $s$ has degree $k_{s}$, defined as the number of edges connected to $s$.
The degree distribution $p_{k}$ is the number of nodes with degree $k$, or
after normalisation the fraction of nodes with degree $k$, or the probability
of a selected node having degree $k$.\cite{Networks}

We are interested in whether the degree distribution follows a power law,\cite{BarabasiA99} such that
\begin{align}
\label{eq:powerLaw}
\begin{array}{rcl}
p_{k} &=& C k^{-\alpha}, \\
\ln p_{k} &=& \ln C - \alpha \ln k,
\end{array}
\end{align}
where $C$ is a constant that is set by the normalisation, and $\alpha$ is to be
determined by the best fit to the data. Clauset \textit{et
al.}\cite{ClausetSN09} have suggested a statistical test based on the method of
maximum likelihood for determining the best fit. The maximum likelihood
estimator (MLE) is\cite{Muniruzzaman57}:
\begin{equation}
\label{eq:mle}
\alpha = 1 + n_{k} \left[\sum\limits_{k_{i} \geq k_{min}} \ln \frac{\displaystyle k_{i}}{\displaystyle k_{min} - \frac{1}{2}}\right]^{-1},
\end{equation}
where $n_{k}$ here is the number of nodes with degree at least $k_{min}$ and
the sum runs over only those nodes. This formula gives an estimate of the best
value of $\alpha$, but gives no indication of the quality of the fit.

Typically, a power law is only obeyed in the tail of the distribution, so a
cutoff $k_{min}$ above which the power law applies needs to be determined.
Clauset \textit{et al.} also have a method for determining the best value of
$k_{min}$.\cite{ClausetYG07} A fit is calculated for all possible values of
$k_{min}$ and the value which gives the smallest Kolmogorov-Smirnov (KS)
statistic \cite{NumericalComputing} is chosen.

A goodness-of-fit test is used to check whether the proposed power law is a
plausible fit for the measured data.\cite{ClausetSN09} Data sets are generated
from the power law and compared to the measured data. The $p$ value is the
fraction of generated data sets that are a worse fit than the measured data.
The fit is rejected if the $p$ value is less than 0.05. If $p > 0.05$, it does
not necessarily indicate that the power law is a good fit, merely that we do
not have sufficient evidence to reject it. 

Other distributions must be considered that could be a better fit to the data,
such as the log-normal distribution. If we cannot say that a power law is a
better fit than the log-normal, we should not conclude that the data follows a
power law.\cite{ClausetSN09} A likelihood ratio test is used to compare the fit
of two distributions, indicating which distribution is more likely, and a $p$
value is used to judge whether the result is significant.\cite{Vuong89} A small $p$
value ($ < 0.05$) means there is sufficient evidence to accept the result of
the tests.

\subsection{Clustering}
Clustering in a network is the degree to which the neighbours of a node are
also neighbours of each other. A local clustering coefficient can be defined
for an individual node $s$ as:\cite{WattsS98}
\begin{equation}
\label{eq:localClustering}
c_s = \frac{\displaystyle \text{number of pairs of neighbours of $s$ that are
connected}}{\displaystyle \text{number of pairs of neighbours of } s}.
\end{equation}
To evaluate the numerator, for each of the nodes present in the adjacency list
of $s$, we count how many of the adjacency lists of the other neighbours contain
the same node. The denominator is easy to evaluate, as
$\frac{1}{2}k_{s}\left(k_s - 1\right)$. The local clustering coefficient is an
indicator of how important a node is in allowing flow through the network in
its local neighbourhood.

The local clustering coefficient can be extended to a global clustering
coefficient for the network in two ways. The average clustering coefficient is
the mean of the local clustering coefficients\cite{WattsS98}:
\begin{equation}
\label{eq:averageClustering}
C_{av} = \frac{\displaystyle 1}{\displaystyle n}\sum_{s} c_{s}.
\end{equation}
The transitivity\cite{LuceP49} depends on the number of triads and triangles in
the network. A triad exists if nodes $r$ and $s$ are neighbours and nodes $s$
and $t$ are neighbours. A triad is a triangle if $r$ and $t$ are also
neighbours. The transitivity is given by:
\begin{equation}
\label{eq:transitivity}
C_{T} = 3 \times \frac{\displaystyle \text{number of triangles}}{\displaystyle \text{number of triads}},
\end{equation}
where the factor of three accounts for the triple counting of triads that are
also triangles. The transitivity can be calculated by extending the calculation
of the local clustering coefficient to look at the neighbours of every node.
The numerator is the number of pairs of neighbours of all nodes that are also
connected. The denominator is $\sum_{s}\frac{1}{2}k_{s}\left(k_s - 1\right)$.
The algorithmic complexity depends on the degree distribution, but in the worst
case is $O\left(n^2\right)$.

These two measures differ in the weighting they give to nodes of small degree.
Such nodes have a small number of possible neighbour pairs, so they have a small
denominator in their local clustering coefficient. They therefore 
dominate the sum in equation \eqref{eq:averageClustering}. In networks
with a large number of small degree nodes, the average clustering coefficient
does not depend much on the larger degree nodes. In PES networks, the large
degree nodes tend to be those with a low potential energy, which are considered
more important for the properties of the network. We therefore prefer the
transitivity to the average clustering coefficient. 

To determine whether or not the transitivity should be considered large or
small, it is useful to compare it to the value for an Erd\H{o}s-R\'{e}nyi
random network\cite{ErdosR59} with the same number of nodes and edges. In this
model, every edge has an equal probability of being present, independent of the
existence of any other edge. Therefore the probability of a given triad being a
triangle, which is the probability of the final edge being present, is the
probability of any edge existing. The maximum number of edges is the number of
possible pairs of nodes, $n \left(n - 1\right)/2$, so given the number of edges
($m$) and nodes ($n$), the transitivity is:
\begin{equation}
\label{eq:randomTransitivity}
C_{Random} = \frac{\displaystyle 2 m}{\displaystyle n \left(n - 1\right)}.
\end{equation}

\subsection{Average Shortest Path Length}
A path in a network is a sequence of steps between nodes along edges. For any
pair of nodes the shortest path (or geodesic path) is defined as the one
connecting the two nodes containing the least number of edges.\cite{Networks}
The shortest path need not be unique, but here we are concerned only with its length,
which is the number of edges it contains. If no path exists between a pair of
nodes, then the network is unconnected. The complete network for a PES must be
connected, as it is impossible to have minima on the surface that cannot be
reached from any other minima through transition states for models with
continuous degrees of freedom. The length of the shortest path between one node
$s$ and all others can be calculated using a breadth-first search.\cite{BFS}

\subsection{Assortativity}
The assortativity\cite{Newman02} of a network is a measure of the tendency of
nodes to have connections with other similar nodes. Specifically, the degree
assortativity refers to whether nodes connect to other nodes of a similar
degree. The assortativity coefficient takes values between 1 and $-1$, where 1
indicates connections between nodes of similar degree, 0 indicates no
preference and $-1$ means that high degree nodes have a preference to connect to
low degree nodes. The definition is,
\begin{equation}
\label{eq:assortativity}
\rho_D = \frac{\displaystyle \sum_{jk} jk \left( e_{jk} - q_j q_k \right)}{\displaystyle \sigma_q^2},
\end{equation}
where $j$ and $k$ are the remaining degrees of two nodes (the degree of the
node excluding the edge connecting the pair), $q_{j}$ is the remaining degree
distribution of $j$, $e_{jk}$ is the joint probability distribution of $q_j$
and $q_k$, and $\sigma_{q}$ is the standard deviation of $q$. The sum runs over
the possible values of $j$ and $k$. This is equivalent to the Pearson
correlation coefficient\cite{Pearson95} between degrees of nodes at the ends of
each edge.

\section{Results}
\label{sec:results}

\begin{figure}[ht]
\centering
    \centering
    \psfrag{ 1}{1}
    \psfrag{ 10}{10}
    \psfrag{ 100}{100}
    \psfrag{ 1000}{1000}
    \psfrag{ 10000}{10000}
    \psfrag{ 100000}{100000}
    \psfrag{ 5}{5}
    \psfrag{ 10}{10}
    \psfrag{ 15}{15}
    \psfrag{ 20}{20}
    \psfrag{ 25}{25}
    \psfrag{ 30}{30}
    \psfrag{rho_3}{$\rho = ~3$}
    \psfrag{rho_6}{$\rho = ~6$}
    \psfrag{rho_10}{$\rho = 10$}
    \psfrag{rho_14}{$\rho = 14$}
    \psfrag{rho_30}{$\rho = 30$}
    \psfrag{Minima and transition states}{Number of minima and transition states}
    \psfrag{Particles}{$N$}
    \includegraphics[scale=1.1]{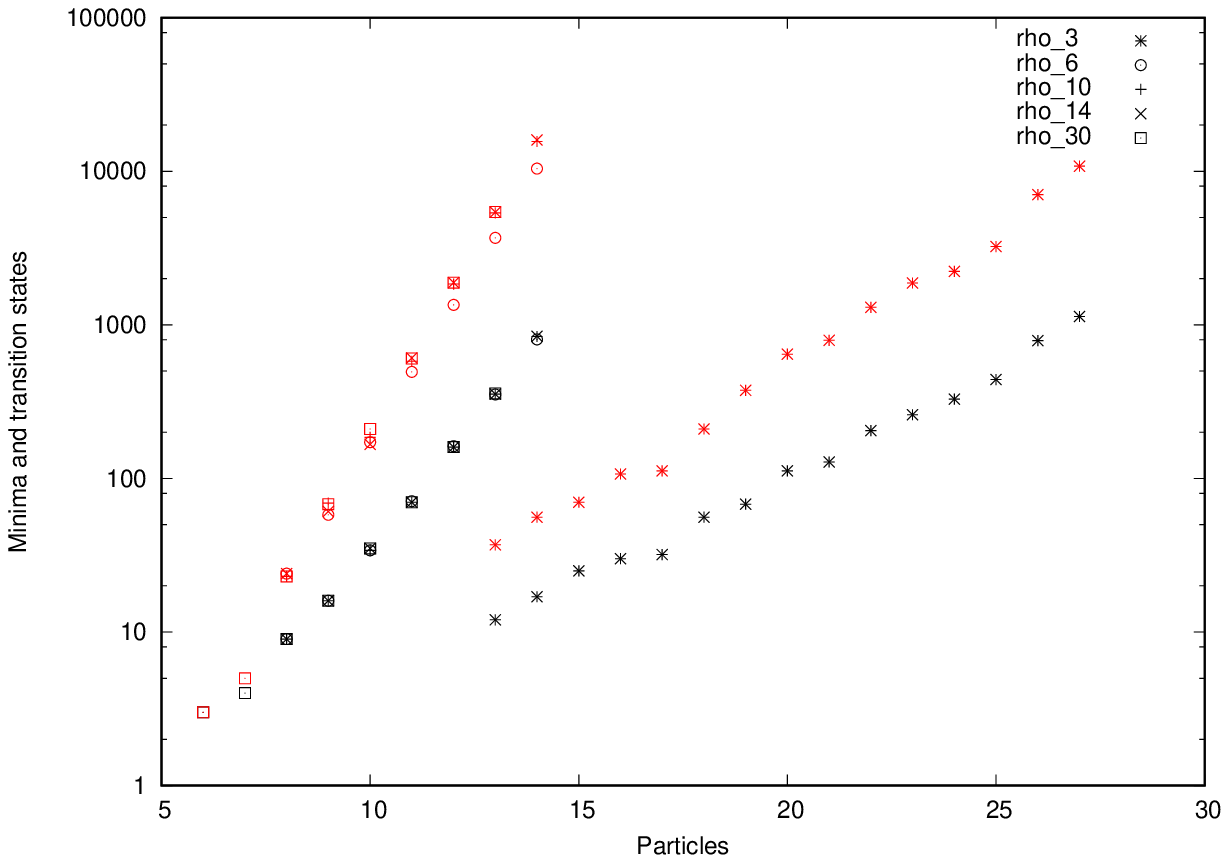}
    \includegraphics[scale=1.1]{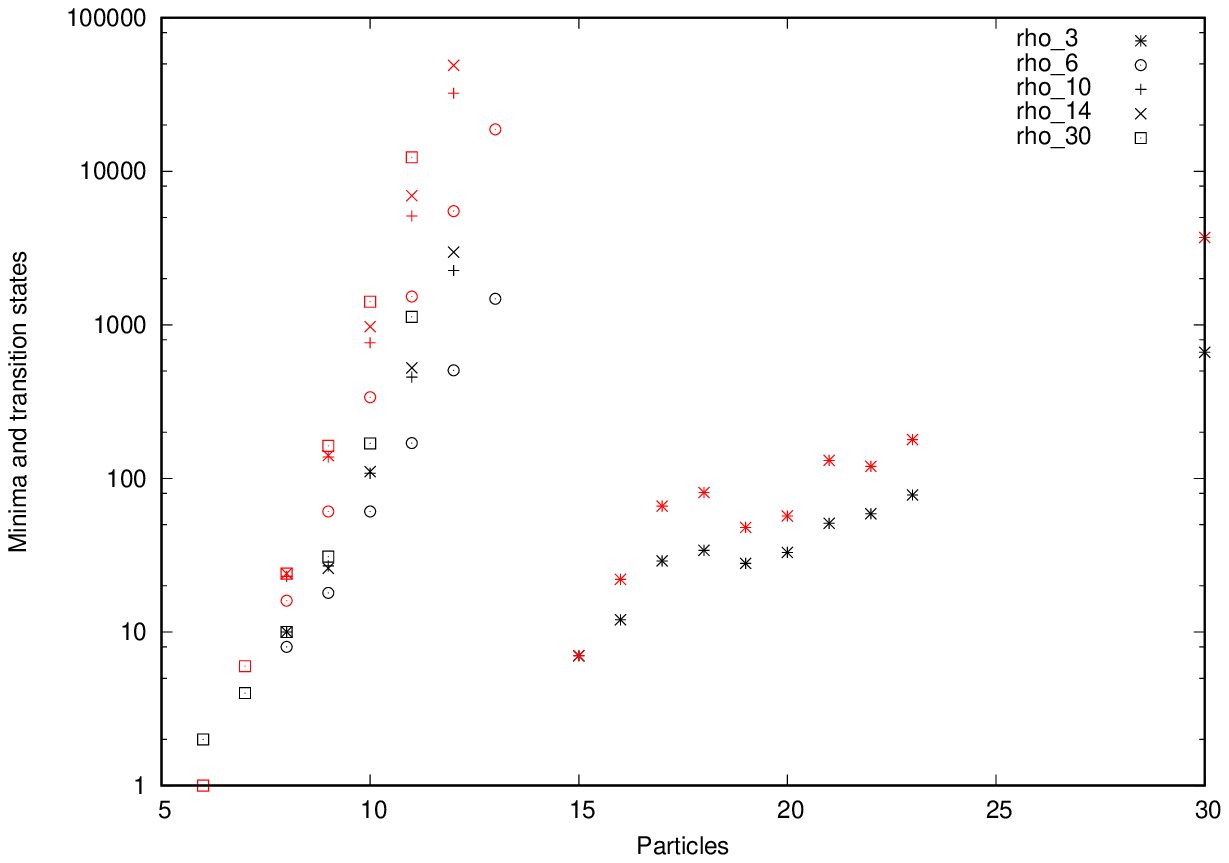}
    \caption{The number or minima (black) and transition states (red) for all
the two-dimensional (top) and three-dimensional (bottom) clusters considered.}
  \label{fig:networkSizes}
\end{figure}

Table \ref{tab:networkSizes} shows the number of minima and transition states found for the largest network
considered at each range in two and three dimensions, after the removal of transition
states corresponding to alternative connections of the same minima and degenerate rearrangements.
Figure \ref{fig:networkSizes} illustrates
the number of minima and transition states for all the clusters considered.

\begin{table}
  \centering
  \begin{tabular}{ccccc}
    \hline
    Dimensions & $\rho$ & $N$ & Minima & Transition States \\ \hline
    2          & 3      & 27  & 1135   & 17006             \\
    2          & 6      & 14  & 805    & 13098             \\
    2          & 10     & 14  & 840    & 19965             \\
    2          & 14     & 14  & 843    & 20552             \\
    2          & 30     & 13  & 358    & 6966              \\
    3          & 3      & 30  & 663    & 6678              \\
    3          & 6      & 13  & 1478   & 25173             \\
    3          & 10     & 12  & 2258   & 41441             \\
    3          & 14     & 12  & 2980   & 60615             \\
    3          & 30     & 11  & 1127   & 13752             \\ \hline
  \end{tabular}
  \caption{The sizes of the largest networks considered for Morse clusters
of $N$ atoms in 2 and 3 dimensions.}
  \label{tab:networkSizes}
\end{table}

\begin{figure}[ht]
\centering
  \begin{minipage}{.45\textwidth}
    \centering
    \psfrag{epsilon}{$\epsilon$}
    \includegraphics[scale=0.45]{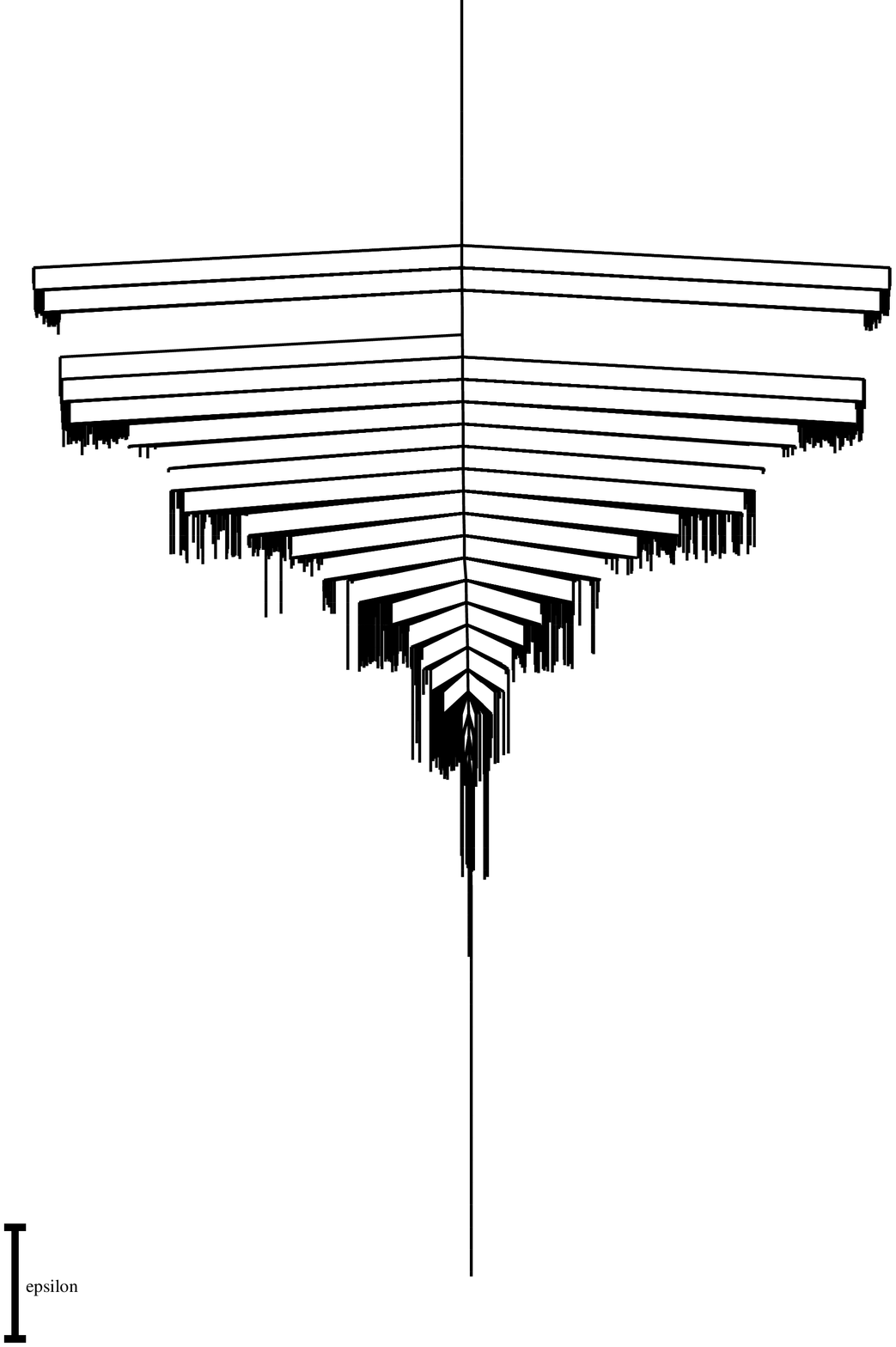} 
  \end{minipage}
  \hspace{0.5cm}
  \begin{minipage}{.45\textwidth}
    \centering
    \psfrag{epsilon}{$\epsilon$}
    \includegraphics[scale=0.45]{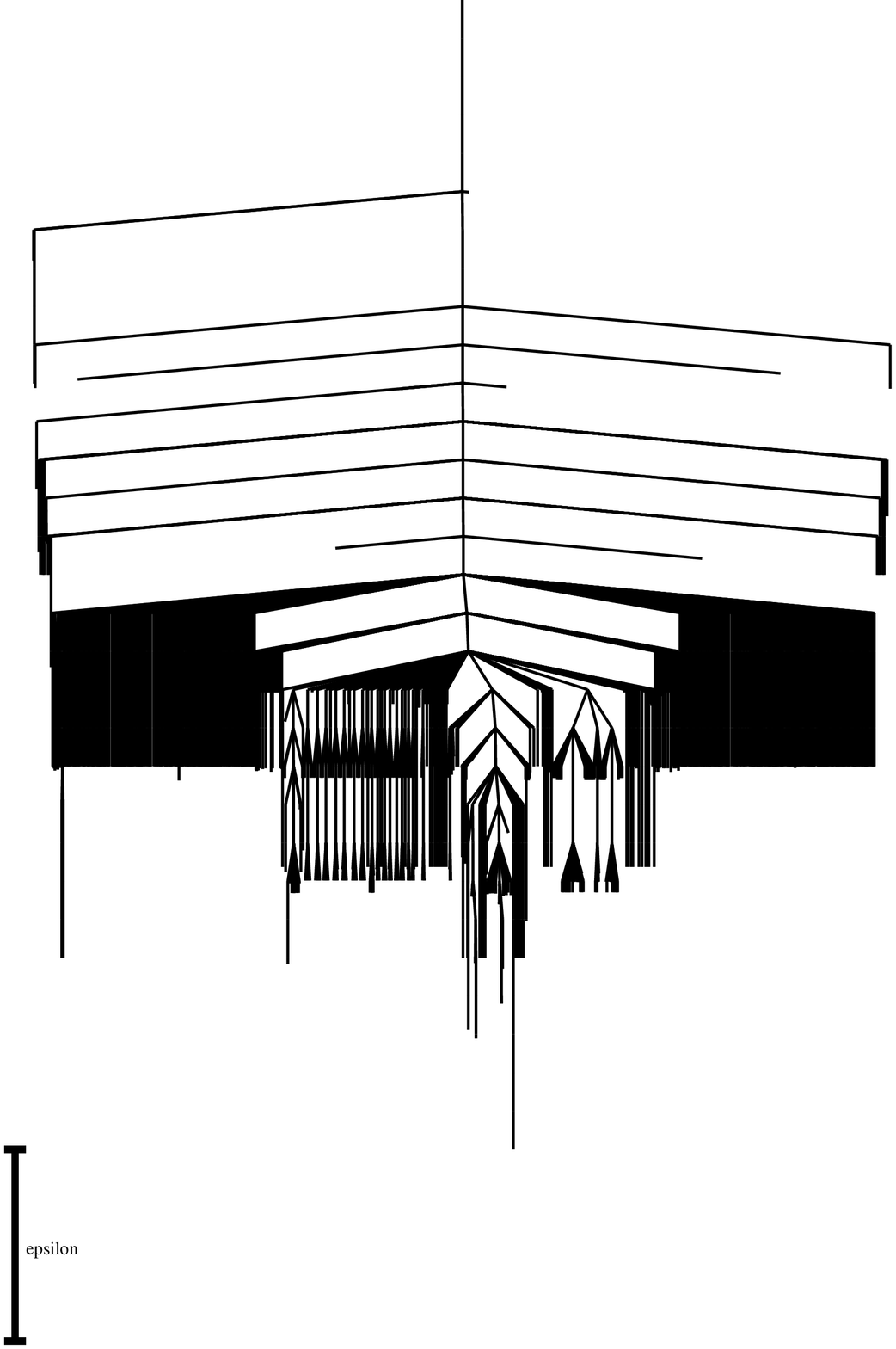} 
  \end{minipage}
  \caption{Disconnectivity graphs for M$^{3D}_{13}$ with $\rho = 6$ (left) and
M$^{3D}_{11}$ with $\rho = 30$ (right). The difference in the global structure
of the landscape is clear.}
  \label{fig:networkDisconnect}
\end{figure}

Some general patterns are clear: 3D clusters have more minima for the same
range and number of particles than 2D clusters, except at very long-range with
$\rho = 3$. The well-known approximate exponential growth of the number of minima and
transition states with system size is present.\cite{StillingerW84, WalesD03}
A shorter-range potential produces an
energy landscape that is locally rough, having more minima, but globally smooth
with a less overall funnelled structure.\cite{CalvoDW12}
To avoid the databases growing too large to be confident of finding the complete network, it was necessary to restrict the number of particles
in the cluster as the range decreased.
The difference in the database size is more apparent in three dimensions than
two dimensions. This effect is evident in the disconnectivity graphs in
figure \ref{fig:networkDisconnect}, where the funnel for a $\rho = 6$
landscape is apparent, suggesting structure-seeking behaviour, but there is no
such structure for $\rho = 30$, where the landscape is more frustrated in the
region of the global minimum.

\subsection{Small-World Behaviour}

\begin{figure}[ht]
  \centering
  \psfrag{ 0.8}{0.8}
  \psfrag{ 1.0}{1.0}
  \psfrag{ 1.2}{1.2}
  \psfrag{ 1.4}{1.4}
  \psfrag{ 1.6}{1.6}
  \psfrag{ 1.8}{1.8}
  \psfrag{ 2.0}{2.0}
  \psfrag{ 2.2}{2.2}
  \psfrag{ 2.4}{2.4}
  \psfrag{ 2.6}{2.6}
  \psfrag{ 2.8}{2.8}
  \psfrag{ 3.0}{3.0}
  \psfrag{ 1}{1}
  \psfrag{ 10}{10}
  \psfrag{ 100}{100}
  \psfrag{ 1000}{1000}
  \psfrag{ 10000}{10000}
  \psfrag{rho = 3}{$\rho = ~3$}
  \psfrag{rho = 6}{$\rho = ~6$}
  \psfrag{rho = 10}{$\rho = 10$}
  \psfrag{rho = 14}{$\rho = 14$}
  \psfrag{rho = 30}{$\rho = 30$}
  \psfrag{Number of minima}{Number of minima}
  \psfrag{Average shortest path length}{Average shortest path length}
  \includegraphics[scale=1.1]{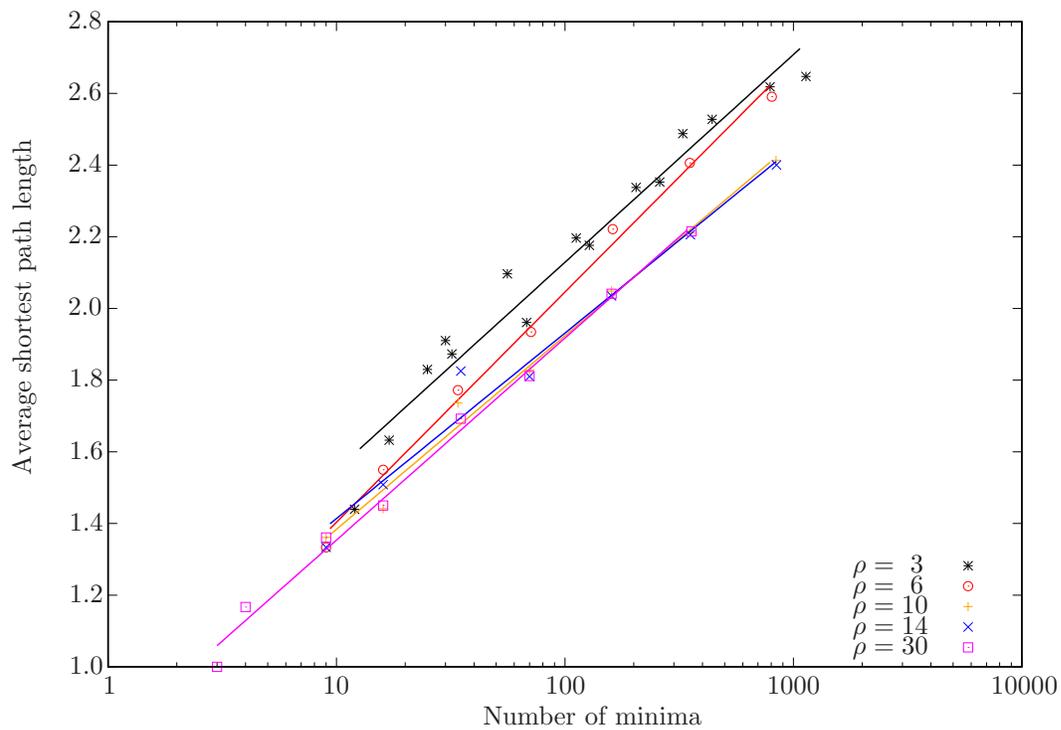}
  \includegraphics[scale=1.1]{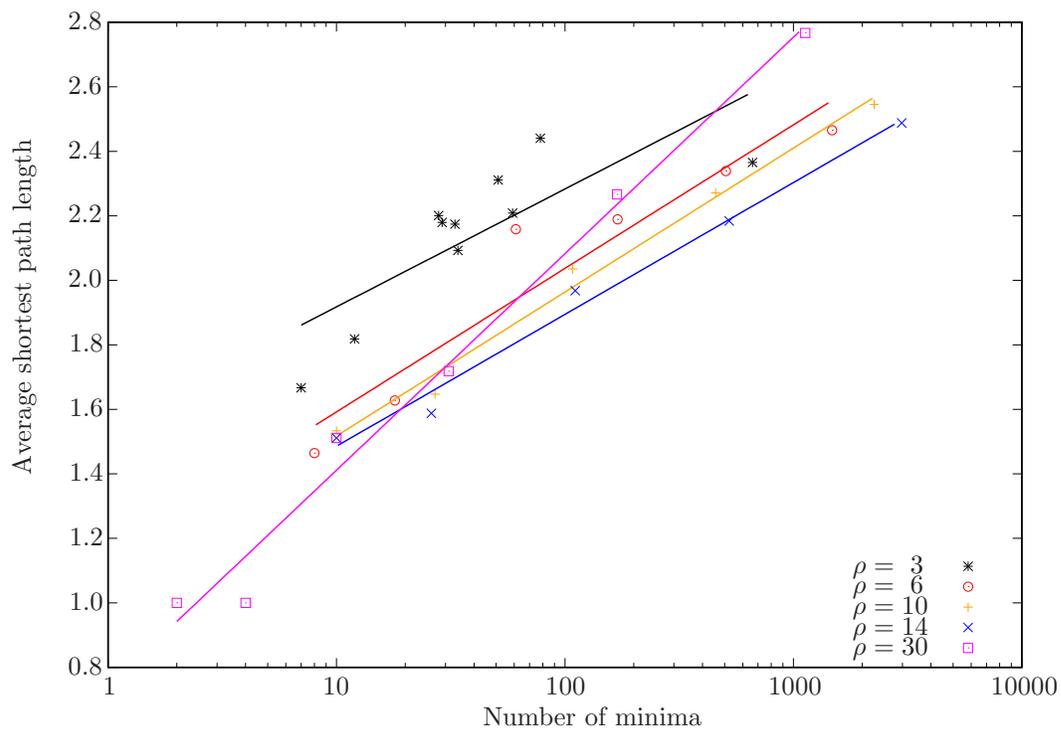}
  \caption{Plots of the average shortest path length against the number of
minima for all the 2D clusters (top) and all the 3D clusters (bottom). Note the
logarithmic scale on the horizontal axis.}
  \label{fig:networkLength}
\end{figure}

To test whether or not the networks show small-world behaviour, the average
shortest path length and the transitivity were considered. The average shortest
path lengths for all the clusters are given in figure \ref{fig:networkLength},
and they show the expected logarithmic dependence on the number of nodes. The
path lengths become slightly smaller as the range decreases. The number of
transition states increases more rapidly with decreasing range than the
number of minima, so a small decrease in the path length with decreasing range
is not surprising.

\begin{figure}[ht]
  \centering
  \psfrag{ 0}{0}
  \psfrag{ 1}{1}
  \psfrag{ 2}{2}
  \psfrag{ 3}{3}
  \psfrag{ 4}{4}
  \psfrag{ 5}{5}
  \psfrag{ 6}{6}
  \psfrag{ 7}{7}
  \psfrag{ 8}{8}
  \psfrag{ 9}{9}
  \psfrag{ 10}{10}
  \psfrag{ 12}{12}
  \psfrag{ 100}{100}
  \psfrag{ 1000}{1000}
  \psfrag{ 10000}{10000}
  \psfrag{Transitivity}{Transitivity / Random transitivity}
  \psfrag{Number of minima}{Number of minima}
  \psfrag{rho = 3}{$\rho = ~3$}
  \psfrag{rho = 6}{$\rho = ~6$}
  \psfrag{rho = 10}{$\rho = 10$}
  \psfrag{rho = 14}{$\rho = 14$}
  \psfrag{rho = 30}{$\rho = 30$}
  \includegraphics[scale=1.1]{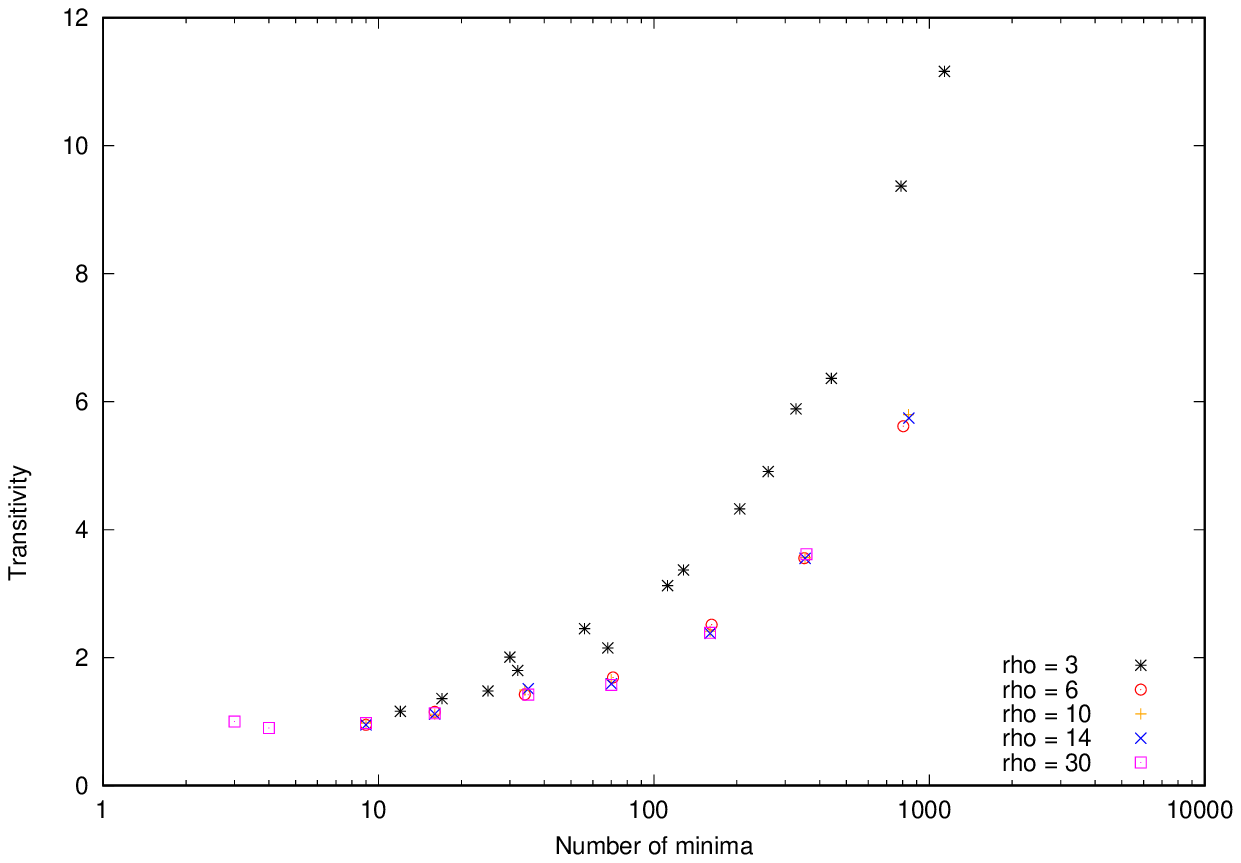}
  \includegraphics[scale=1.1]{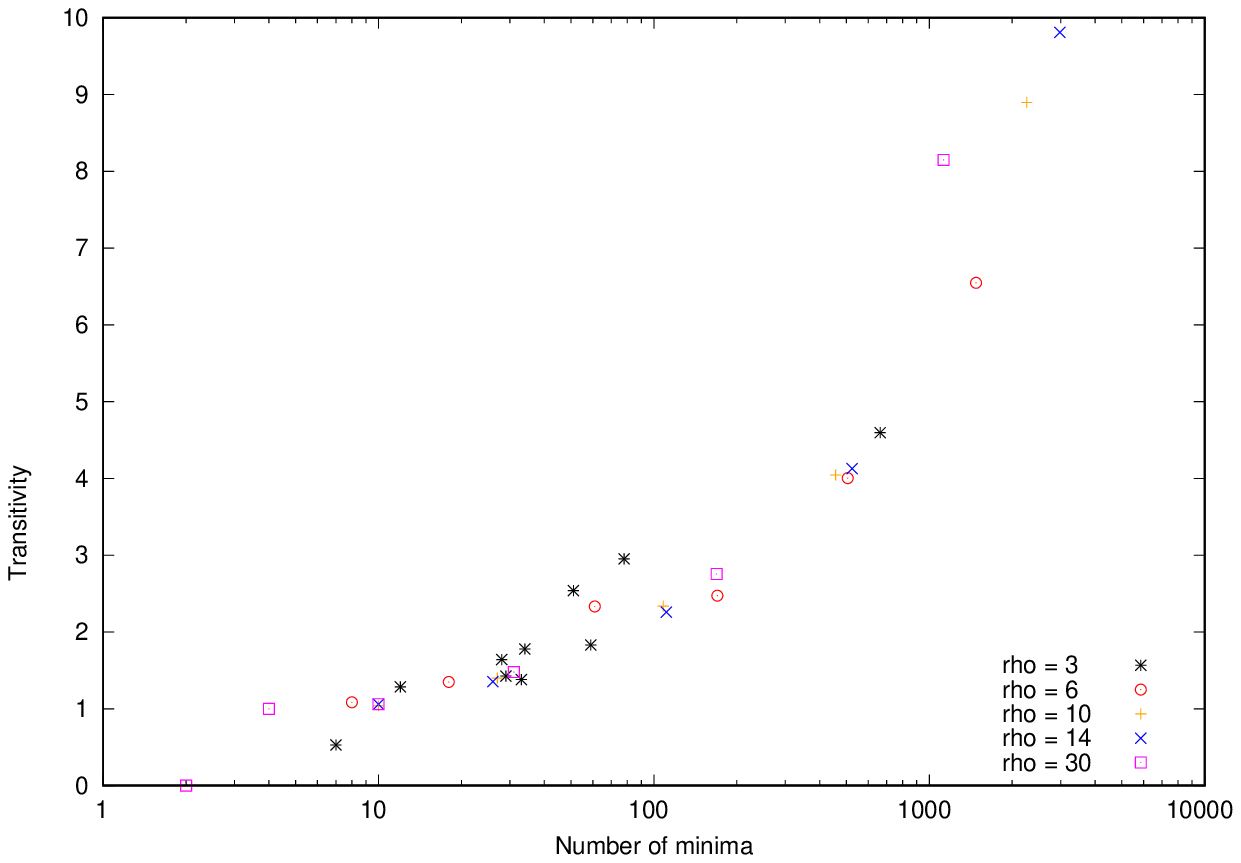}
  \caption{Plots of the ratio of the transitivity of each network to the
transitivity of a random graph with the same number of nodes and edges, against
the number of minima for all the 2D clusters (top) and all the 3D clusters
(bottom).}
  \label{fig:networkClustering}
\end{figure}

The transitivity for a small-world network is expected to be larger than for the
equivalent random graph with the same number of minima and
edges.\cite{WattsS98} The transitivity scaled by the value for a random graph
with the same number of nodes and edges is plotted for all the clusters in
figure \ref{fig:networkClustering}. The expected small-world behaviour is again
apparent, with the transitivity larger than for a random graph in all
the clusters, except for three-dimensional systems with very few minima,
where finite size effects dominate. There is no significant difference between
the values for different ranges, except for $\rho = 3$, which appears to have a
slightly higher transitivity. The long range of the potential produces a smoother
landscape, which makes it more likely that minima in distant parts of
configuration space are directly connected.
Short-range clusters do not display the structure-seeking
properties of long-range clusters, since their landscapes are not funnelled,
and there are relatively large barriers separating some minima with similar energies. 
Since the present analysis
accounts only for the existence of connections, regardless of
barrier heights, we find that the small-world
properties of the network are similar for short- and long-range clusters.

\begin{figure}[ht]
  \centering
  \psfrag{-43}{}
  \psfrag{-42}{$-42$}
  \psfrag{-41}{}
  \psfrag{-40}{$-40$}
  \psfrag{-39}{}
  \psfrag{-38}{$-38$}
  \psfrag{-37}{}
  \psfrag{-36}{$-36$}
  \psfrag{-35}{}
  \psfrag{-34}{$-34$}
  \psfrag{-33}{}
  \psfrag{-29.5}{}
  \psfrag{-29}{$-29$}
  \psfrag{-28.5}{}
  \psfrag{-28}{$-28$}
  \psfrag{-27.5}{}
  \psfrag{-27}{$-27$}
  \psfrag{-26.5}{}
  \psfrag{-26}{$-26$}
  \psfrag{-25.5}{}
  \psfrag{-25}{$-25$}
  \psfrag{-24.5}{}
  \psfrag{-24}{$-24$}
  \psfrag{ 1}{1}
  \psfrag{ 10}{10}
  \psfrag{ 100}{100}
  \psfrag{ 1000}{1000}
  \psfrag{Degree}{Degree}
  \psfrag{Potential Energy}{Potential Energy}
  \includegraphics[scale=1.1]{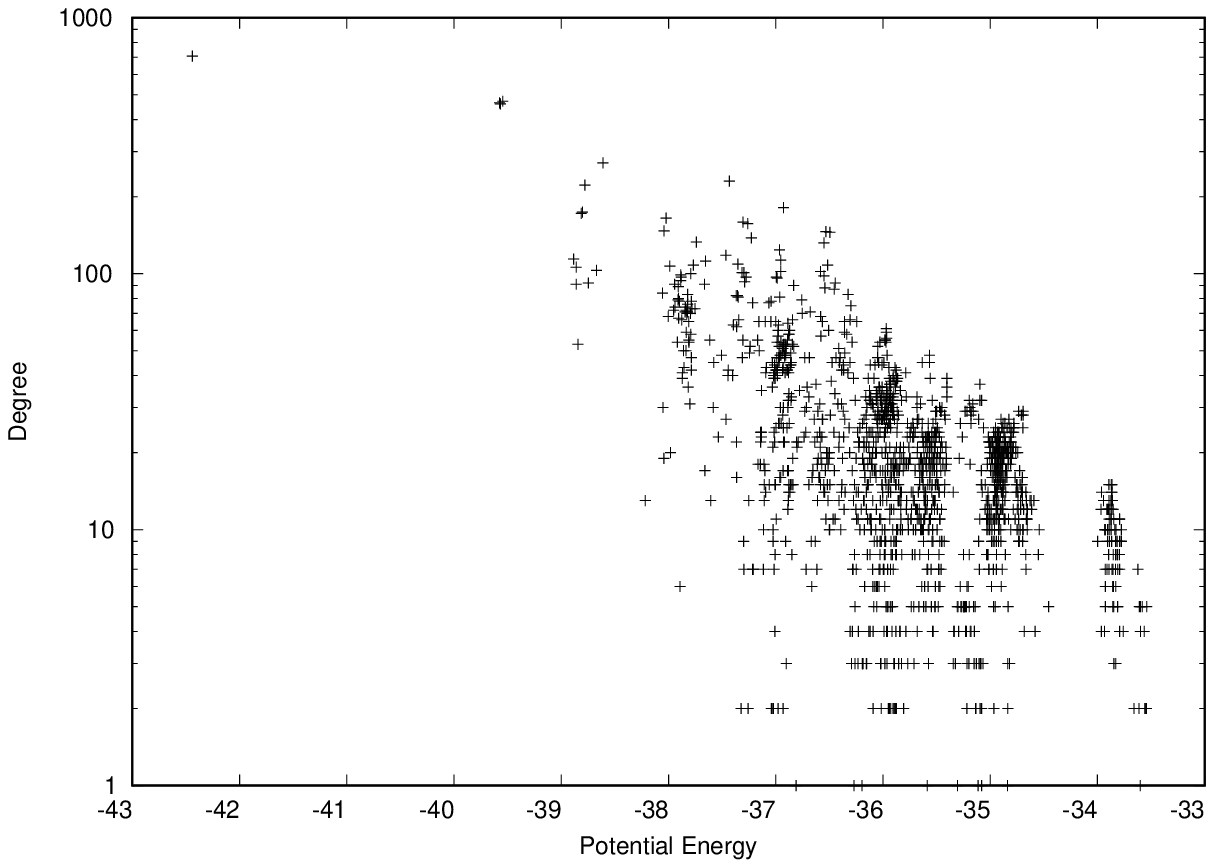}
  \includegraphics[scale=1.1]{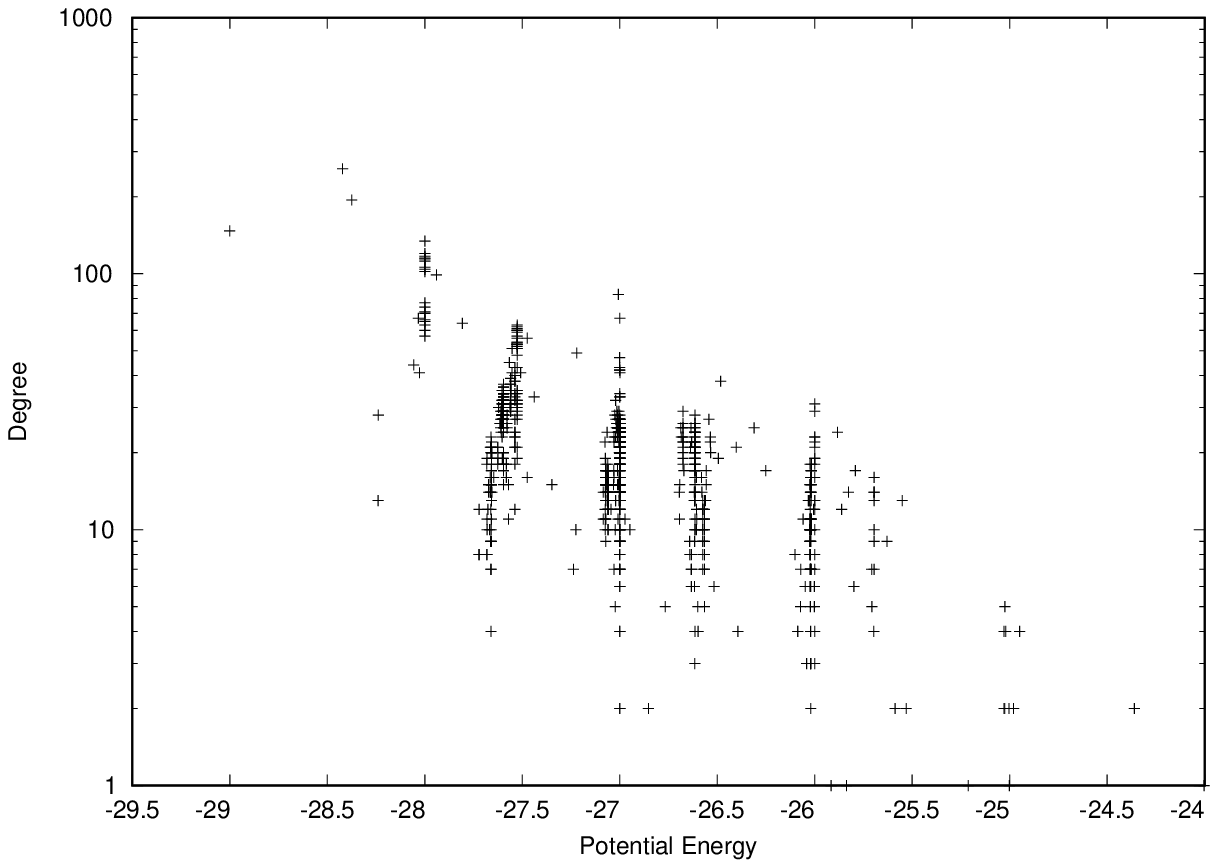}
  \caption{Degree of each minimum as a function of potential energy for
M$^{3D}_{13}$ with $\rho = 6$ (top) and M$^{3D}_{11}$ with $\rho = 30$
(bottom)}
  \label{fig:networkPot}
\end{figure}

Insight into the structure of the landscape and network can be obtained by
inspecting a plot of the degree of the nodes against their potential energy, shown
in figure \ref{fig:networkPot} for M$^{3D}_{13}$ with $\rho = 6$ and the
M$^{3D}_{11}$ with $\rho = 30$. The funnelled landscape for the $\rho = 6$
cluster has one node with a significantly lower potential energy, as seen in
the disconnectivity graph (figure \ref{fig:networkDisconnect}), which is also
the node with the highest degree. Over a third of the minima in the network are
directly connected to the global minimum. At higher energies, although there
are many minima with different degrees at similar energies, the maximum degree
for a given energy decreases as the energy increases. These observations can be
related to funnels on the landscape: the global minimum has a large basin of
attraction, which occupies a significant volume in configuration space, giving
a large boundary surface containing many transition states. High energy minima
have small basins of attraction with a small boundary surface and fewer
transition states. In comparison, the global minimum for M$^{3D}_{11}$ with
$\rho = 30$ is not the minimum with the highest degree, and the most highly
connected minimum has transition states linking it to less than a quarter of
the other minima. The general trend for higher energy minima to have a lower
degree is still present. The prominent vertical stripes are a result of the
short-ranged potential. Since the potential energy is primarily determined by
the number of nearest-neighbour contacts, an integer, it
is approximately
discretised.\cite{Wales94b,DoyeW96b,DoyeDW97,BranzMEM02,CalvoBGH05}

\subsection{Scale-Free Behaviour}

To test whether a network is scale-free, we must determine whether the degree
distribution (or cumulative degree distribution) follows a power law, at least
in the high degree range. The fitted $\alpha$ and $k_{min}$ parameters [see
equation \eqref{eq:powerLaw}] and $p$ values from the goodness-of-fit tests for
the largest cluster for each dimension and range parameter are shown in table
\ref{tab:pValue}. We also show $n_{tail}$, the number of minima remaining
after those with degree less than $k_{min}$ have been removed. If $p < 0.05$,
there is good evidence to reject the hypothesis that a power law is a good fit
to the data; otherwise the power law may be a good fit. A large value of
$k_{min}$ and a small value of $n_{tail}$ indicates the fit is only applicable
for a small range of the data and is unreliable.

\begin{table}
  \centering
  \begin{tabular}{ccccccc}
    \hline
    Dimensions & $\rho$ & N  & $\alpha$        & $k_{min}$ & $n_{tail}$ & $p$ \\ \hline
    2          & 3      & 27 & $2.32 \pm 0.32$ & $23 \pm 12$ & 255 & 0.00 \\
    2          & 6      & 14 & $3.50 \pm 0.16$ & $26 \pm 4$  & 302 & 0.00 \\
    2          & 10     & 14 & $2.55 \pm 0.14$ & $18 \pm 7$  & 765 & 0.00 \\
    2          & 14     & 14 & $2.03 \pm 0.21$ & $10 \pm 5$  & 840 & 0.00 \\
    2          & 30     & 13 & $2.90 \pm 0.15$ & $17 \pm 4$  & 332 & 0.00 \\
    3          & 3      & 30 & $2.79 \pm 0.17$ & $10 \pm 2$  & 219 & 0.18 \\
    3          & 6      & 13 & $2.66 \pm 0.16$ & $19 \pm 8$  & 698 & 0.02 \\
    3          & 10     & 12 & $2.56 \pm 0.09$ & $28 \pm 3$  & 498 & 0.03 \\
    3          & 14     & 12 & $2.37 \pm 0.12$ & $65 \pm 11$ & 174 & 0.02 \\
    3          & 30     & 11 & $3.30 \pm 0.25$ & $48 \pm 8$  & 47  & 0.16 \\ \hline
    3          & LJ     & 14 & $2.82 \pm 0.10$ & $37 \pm 8$  & 513 & 0.07 \\ \hline
  \end{tabular}
  \caption{Fitted values of $\alpha$, $k_{min}$ and $p$ values from the
goodness-of-fit tests for the largest Morse cluster considered at each range
and dimension, as well as LJ$_{14}$. If $p < 0.05$, the fit should be
considered poor.}
  \label{tab:pValue}
\end{table}

\begin{figure}[ht]
  \centering
  \psfrag{degree}{Degree}
  \psfrag{frequency}{Frequency}
  \psfrag{10_4}{$10^{-4}$}
  \psfrag{10_3}{$10^{-3}$}
  \psfrag{10_2}{$10^{-2}$}
  \psfrag{10_1}{$10^{-1}$}
  \psfrag{10_0}{$10^{0}$}
  \psfrag{0}{}
  \psfrag{-1}{}
  \psfrag{-2}{}
  \psfrag{-3}{}
  \psfrag{-4}{}
  \psfrag{-5}{}
  \psfrag{1}{1}
  \psfrag{10}{10}
  \psfrag{100}{100}
  \psfrag{1000}{1000}
  \psfrag{foo}{}
  \includegraphics[scale=0.7]{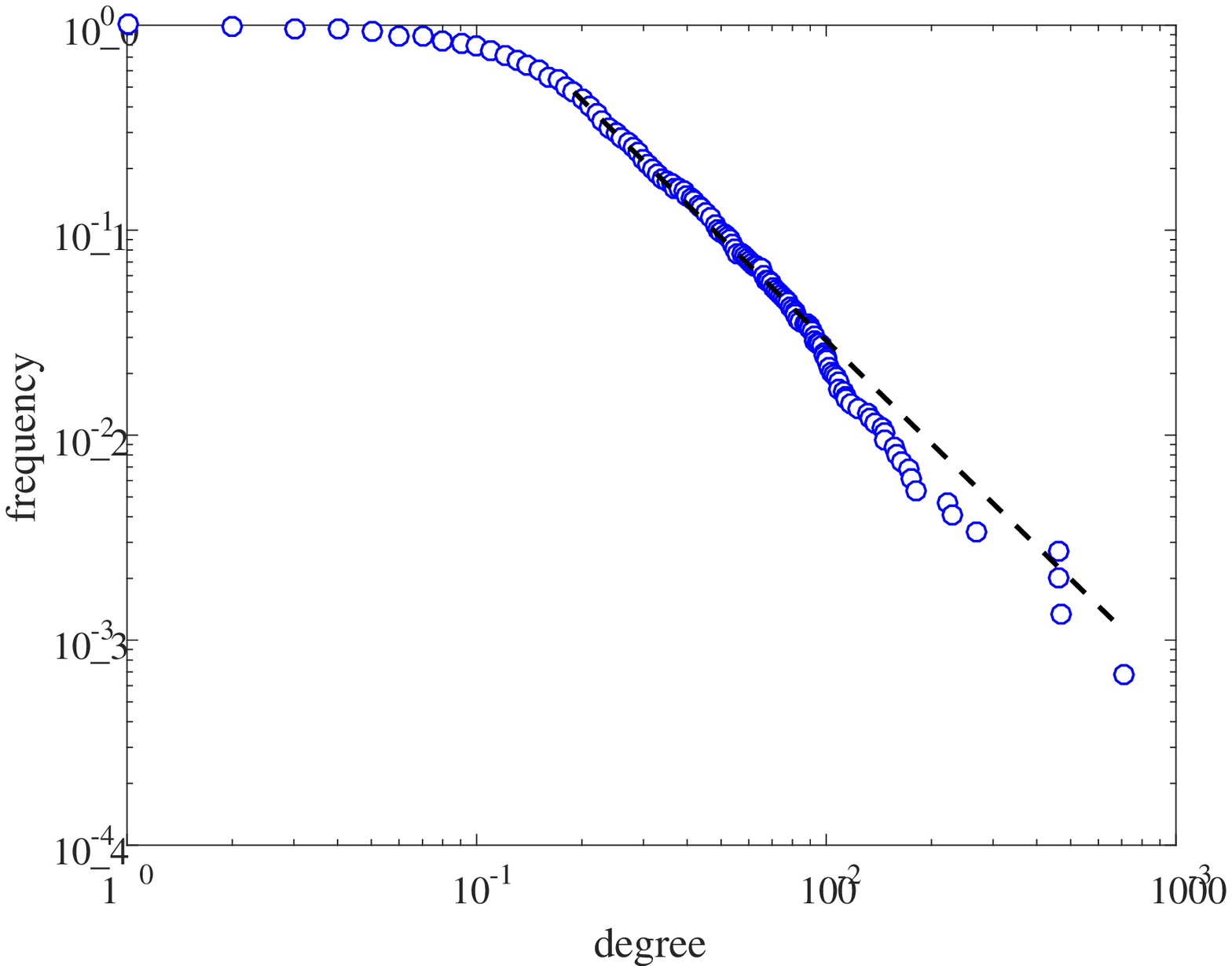}
  \includegraphics[scale=0.7]{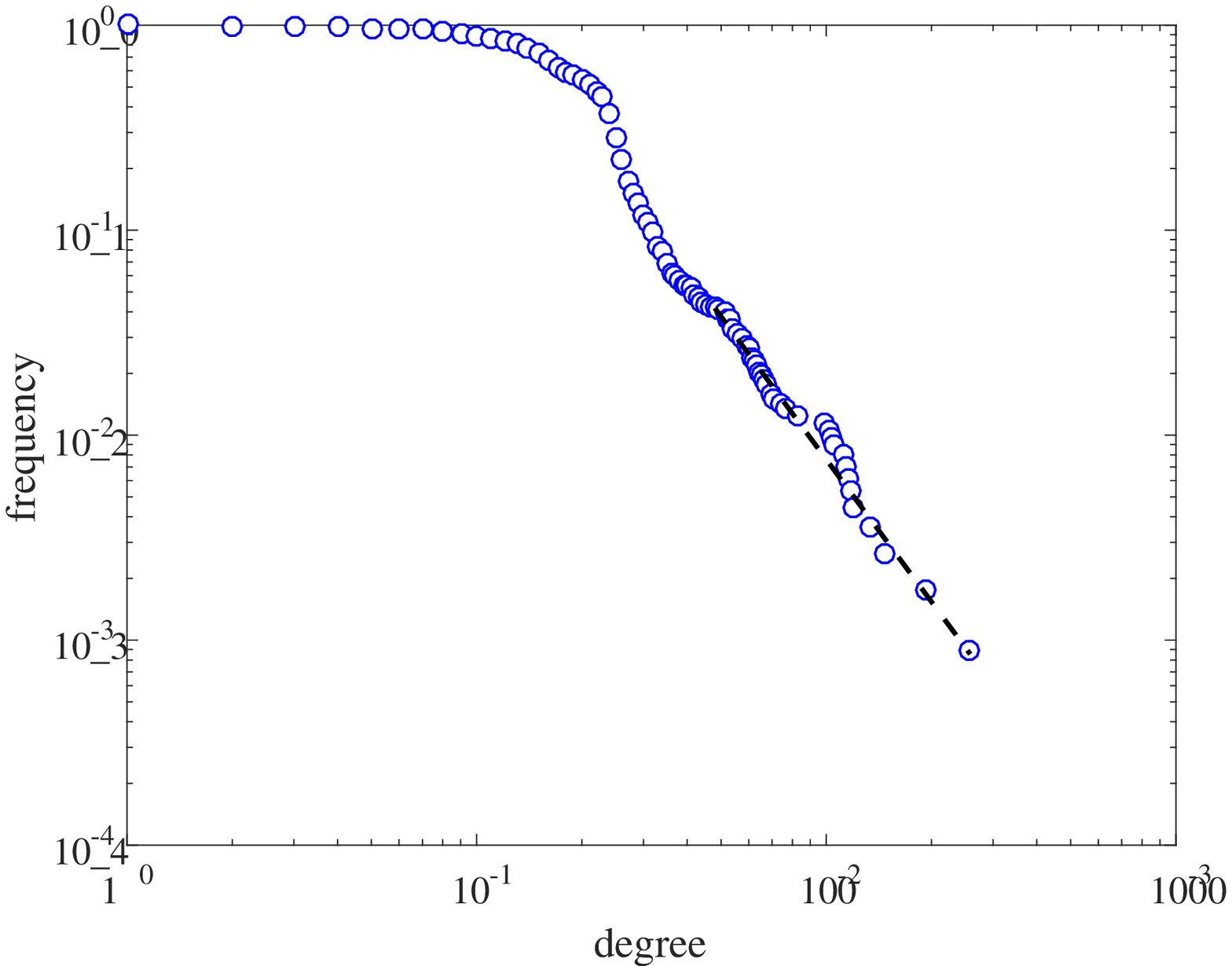}
  \caption{Plots of the normalised cumulative degree distribution for
M$^{3D}_{13}$ with $\rho = 6$ (top) and M$^{3D}_{11}$ with $\rho = 30$
(bottom). The calculated best fit is indicated by the black dashed line.}
  \label{fig:networkDistro}
\end{figure}

The cumulative degree distribution is plotted in figure \ref{fig:networkDistro}
for M$^{3D}_{13}$ clusters 
with $\rho = 6$ (top) and M$^{3D}_{11}$ with $\rho = 30$
(bottom), along with the best power law fit calculated by the MLE method. The
low degree end does not follow a power law, because the number of minima for a
cluster is finite. Above $k_{min}$, the distribution appears to be
approximately linear. However, the goodness-of-fit test for the $\rho = 6$
cluster gives $p < 0.05$, indicating the power law is not a good fit for the
data.

In fact, according to the suggested criteria, only two clusters have a $p$ value compatible
of a power law fit: M$^{3D}_{30}$ with $\rho = 3$ and M$^{3D}_{11}$
with $\rho = 30$. Both of these clusters have a large value of $k_{\min}$
compared to the number of minima in the cluster, giving a small value of
$n_{tail}$. Therefore the fit should be considered unreliable, especially as it
is clearly the case that these clusters do not generally display scale-free
behaviour.

\subsection{Lennard-Jones Clusters}
\begin{figure}[ht]
  \centering
  \psfrag{Degree}{Degree}
  \psfrag{Probability}{Frequency}
  \psfrag{10_5}{$10^{-5}$}
  \psfrag{10_4}{$10^{-4}$}
  \psfrag{10_3}{$10^{-3}$}
  \psfrag{10_2}{$10^{-2}$}
  \psfrag{10_1}{$10^{-1}$}
  \psfrag{10_0}{$10^{0}$}
  \psfrag{-1}{}
  \psfrag{-2}{}
  \psfrag{-3}{}
  \psfrag{-4}{}
  \psfrag{-5}{}
  \psfrag{1}{1}
  \psfrag{10}{10}
  \psfrag{100}{100}
  \psfrag{1000}{1000}
  \psfrag{10000}{10000}
  \psfrag{foo}{}
  \includegraphics[scale=0.8]{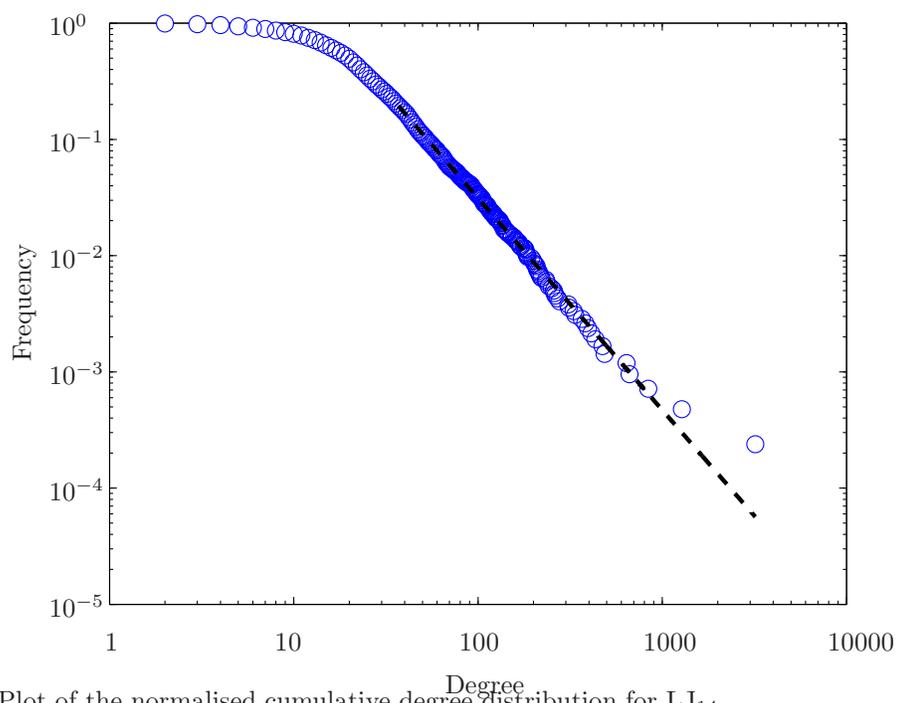}
  \caption{Plot of the normalised cumulative degree distribution for $\mathrm{LJ_{14}}$.}
  \label{fig:networkLJ}
\end{figure}

Doye and Massen\cite{DoyeM05} studied Lennard-Jones (LJ) clusters and concluded that the degree
distribution follows a power law and that the networks show scale-free
behaviour. However, they did not give a statistical analysis, as the appropriate
tools had not been identified at that time, simply stating that `as the size of the clusters
increase a clear power law tail develops.' Using their data for the cluster
with 14 particles,\cite{LJData} which includes 4196 minima and 61085 transition
states, we have now applied the goodness-of-fit tests, as for the Morse
clusters. These data are shown in table \ref{tab:pValue} and the cumulative
degree distribution and fit are shown in figure \ref{fig:networkLJ}. Since $p >
0.05$, this distribution is a candidate for a power law fit. However, we tested
the power law fit against a log-normal fit. The log-likelihood ratio $R =
0.80$, but $p = 0.41$, so the test is inconclusive. Clauset \textit{et al.}
comment that deciding between a power law fit and a log-normal fit can be
problematic.\cite{ClausetSN09} We conclude that the distribution
approximately follows a power law.

\subsection{Bulk Ortho-Terphenyl}
A database of minima and transition states for a bulk representation of the
molecular glass-former ortho-terphenyl (OTP) was provided by Niblett and
coworkers.\cite{NiblettSSW16} The details of the model are described in their
paper. The database was relatively large, containing 313,651 minima
and 334,272 transition states. After removing alternative transition states linking the
same pairs of minima and self-connections (degenerate rearrangements),
332,774 transition states remained. In contrast to the Morse and Lennard-Jones
clusters considered above, the database is certainly very far from
complete, since finding all the minima and transition states for such a glassy
landscape is unfeasible.

Testing whether the incomplete OTP database is likely to be representative of the full
network was assessed by taking increasingly large subsets of the database and
looking at the convergence of the network properties. Here the appropriate
network properties are the ones that are global, but do not directly depend on
the size of the network. The average shortest path length is therefore
unsuitable, as it is expected to increase with the network size. Three
properties were chosen: the transitivity, the average clustering coefficient
(the mean of the local clustering coefficients) and the assortativity.

\begin{figure}[ht]
  \centering
  \psfrag{Number of transition states}{Number of transition states}
  \psfrag{Clustering coefficient}{Clustering coefficient}
  \psfrag{Assortativity}{Assortativity}
  \psfrag{ 0}{0.00}
  \psfrag{ 0.02}{}
  \psfrag{ 0.04}{0.04}
  \psfrag{ 0.06}{}
  \psfrag{ 0.08}{0.08}
  \psfrag{ 0.1}{}
  \psfrag{ 0.12}{0.12}
  \psfrag{ 0.14}{}
  \psfrag{ 0.16}{0.16}
  \psfrag{ 0.18}{}
  \psfrag{ 0.2}{0.20}
  \psfrag{  0.2}{0.2}
  \psfrag{  0}{0.0}
  \psfrag{-0.2}{$-0.2$}
  \psfrag{-0.4}{$-0.4$}
  \psfrag{-0.6}{$-0.6$}
  \psfrag{-0.8}{$-0.8$}
  \psfrag{-1}{$-1.0$}
  \psfrag{ 1}{1}
  \psfrag{ 10}{10}
  \psfrag{ 100}{$10^2$}
  \psfrag{ 1000}{$10^3$}
  \psfrag{ 10000}{$10^4$}
  \psfrag{ 100000}{$10^5$}
  \psfrag{ 1000000}{$10^6$}
  \includegraphics[scale=0.8]{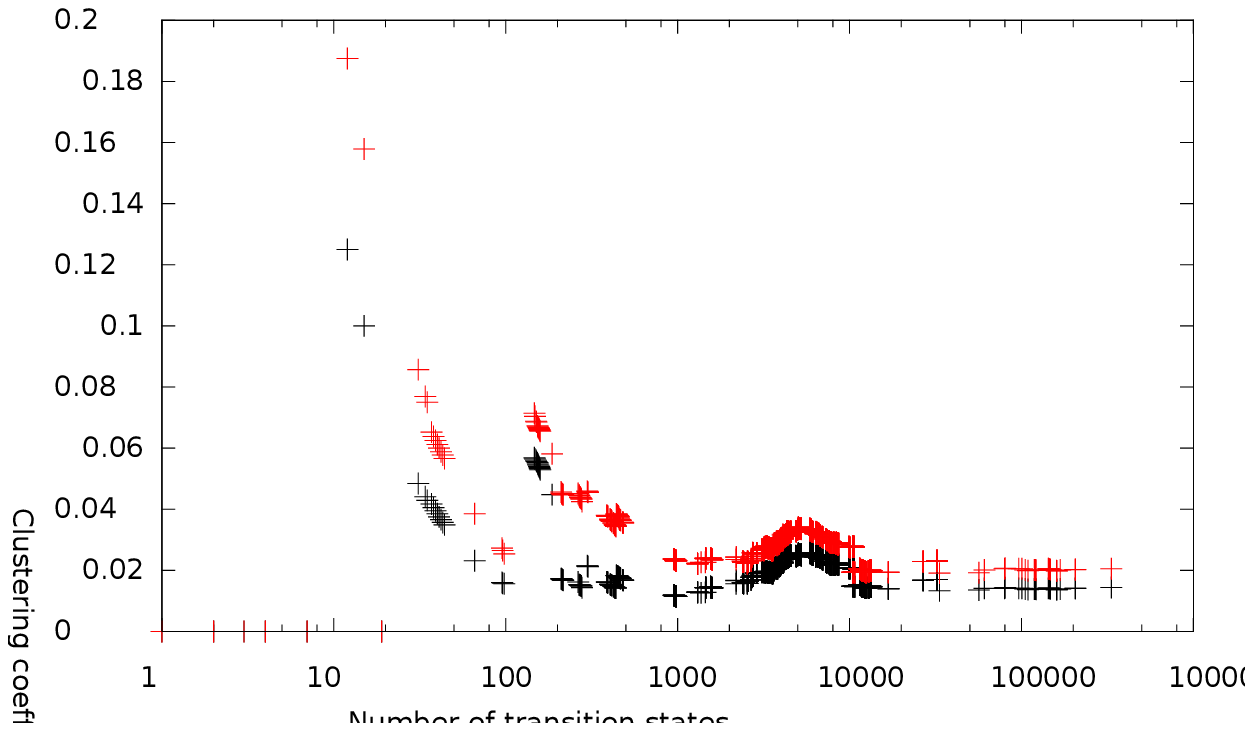}
  \includegraphics[scale=0.8]{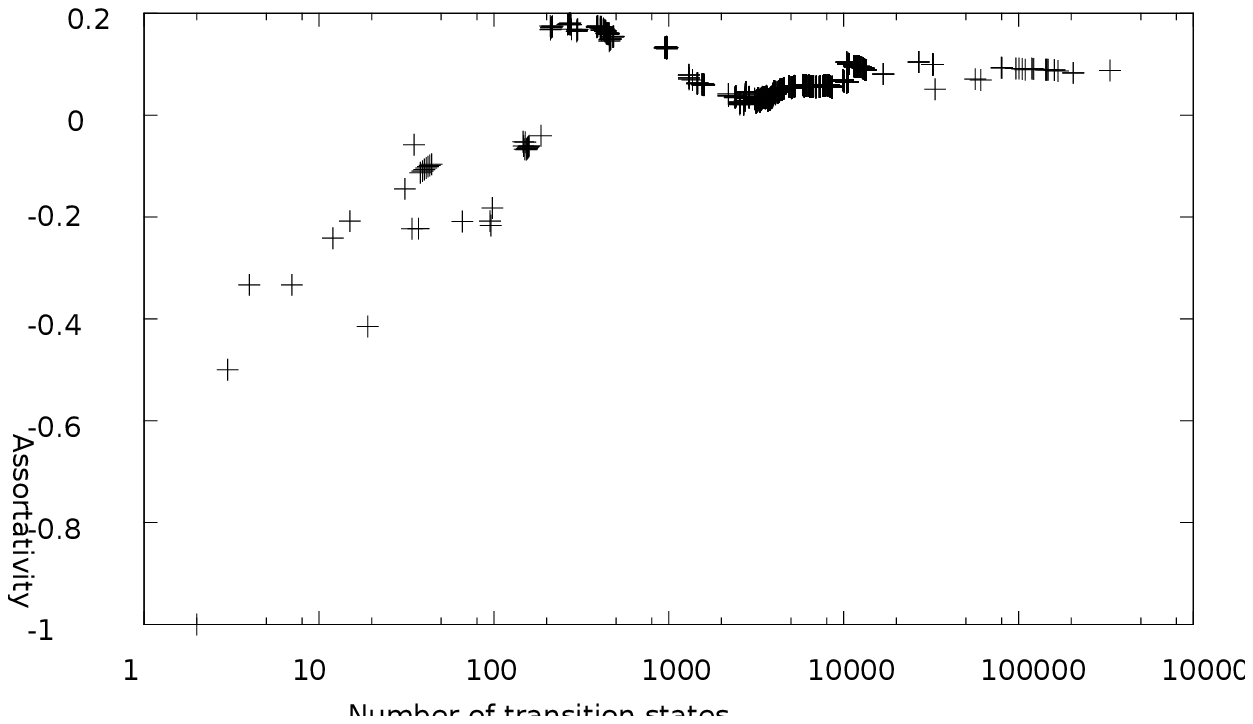}
  \caption{Top: the variation of the average clustering coefficient (black) and
transitivity (red) as the largest connected component of the OTP database
grows. Bottom: the corresponding plot for the assortativity.}
  \label{fig:convergenceOTP}
\end{figure}

All three properties are only defined for a connected network, but an arbitrary
subset of the database will not generally be connected because of the way it
was constructed. Hence transition states were
added to the network one by one, in the order located by the original search,
and the largest connected component selected. Whenever the largest connected
component increased in size, the network properties were calculated. Graphs of
the three properties against the number of transition states in the connected
component are shown in figure \ref{fig:convergenceOTP}. There are fluctuations
in the values, but all three seem to be approaching a limit. However, caution
is needed in interpreting these results. Sampling biases in
generating the database could potentially lead to limiting values of these
properties without truly being a representative sample of the network.

\begin{figure}[ht]
  \centering
  \psfrag{Number of transition states}{Number of transition states}
  \psfrag{Ratio}{Ratio}
  \psfrag{ 0.1}{$10^{-1}$}
  \psfrag{ 1}{1}
  \psfrag{ 10}{10}
  \psfrag{ 100}{$10^2$}
  \psfrag{ 1000}{$10^3$}
  \psfrag{ 10000}{$10^4$}
  \psfrag{ 100000}{$10^5$}
  \psfrag{ 1000000}{$10^6$}
  \includegraphics[scale=0.8]{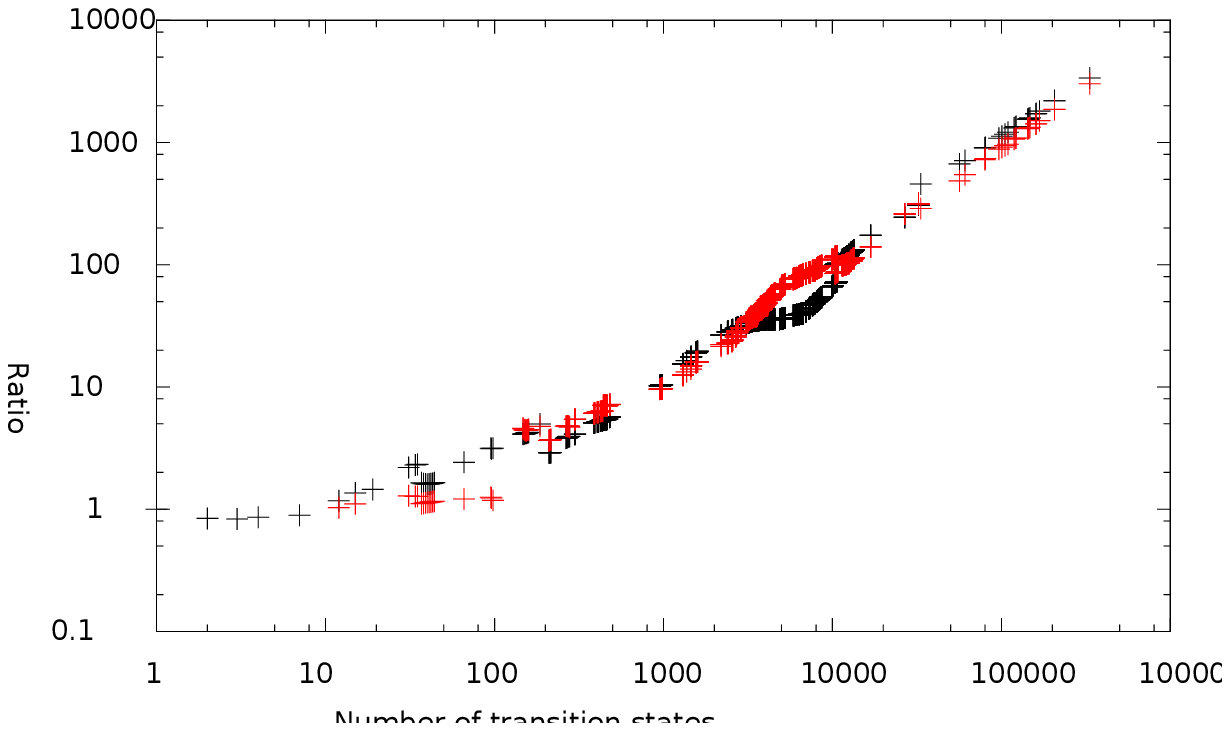}
  \caption{The ratio of average shortest path length for the largest connected
component of the OTP database to that of the equivalent random graph (black)
and the corresponding ratio for the transitivity (red).}
  \label{fig:smallWorldOTP}
\end{figure}

To test for small-world properties, the values of the average shortest path
length and transitivity can be compared to those of a random graph. These
ratios are plotted in figure \ref{fig:smallWorldOTP} for increasingly large
connected subgraphs of the network. The transitivity and average
shortest path length increase more quickly than the values for the equivalent
random graph suggesting that the network (or at least this subset of the full PES)
exhibits lattice graph properties, rather than small-world behaviour. The
network is mildly assortative, although the value is close to zero, indicating
a slight preference for minima to connect to other minima with similar degree.
Assortative networks tend to have cores of highly connected nodes with a
periphery of low degree nodes.\cite{Networks} This pattern could arise in a
glassy landscape, with well connected groups of structurally closely related
minima, and higher energy minima forming the periphery between the numerous
groups.

Scale-free networks are small-world, so this network is also not scale-free. We
found $p = 0.0$ for the power law goodness-of-fit test.

\subsection{Bulk Binary Lennard-Jones}

A database for a bulk BLJ system was provided by de Souza and
Wales.\cite{deSouzaW06,deSouzaW08} The system contained 60 atoms, 48 of type A
and 12 of type B, with interaction parameters $\sigma_{\rm AA} = 1.0,
\sigma_{\rm AB} = 0.8, \sigma_{\rm BB} = 0.88, \epsilon_{\rm AA} = 1.0,
\epsilon_{\rm AB} = 1.5$ and $\epsilon_{\rm BB} = 0.5$. These parameters are
known to give rise to a glassy landscape.\cite{KobA94, KobA95} Periodic
boundary conditions were used with a cubic box of length $3.587$. The database
was connected and contained 11,538 minima and 13,088 transition states,
dropping to 12,957 transition states after removing alternative
transition states linking the same pairs of minima and degenerate rearrangements.

The database was expanded until it contained 703,827 minima and 360,836
transition states. However, the largest connected component consisted of only
12,905 minima and 14,363 transition states.

\begin{figure}[ht]
  \centering
  \psfrag{Number of transition states}{Number of transition states}
  \psfrag{Clustering coefficient}{Clustering coefficient}
  \psfrag{Assortativity}{Assortativity}
  \psfrag{ 0}{0.00}
  \psfrag{ 0.05}{0.05}
  \psfrag{ 0.1}{0.20}
  \psfrag{ 0.15}{0.15}
  \psfrag{ 0.2}{0.20}
  \psfrag{  0.2}{0.2}
  \psfrag{ 0.25}{0.25}
  \psfrag{  0}{0.0}
  \psfrag{-0.2}{$-0.2$}
  \psfrag{-0.4}{$-0.4$}
  \psfrag{-0.6}{$-0.6$}
  \psfrag{-0.8}{$-0.8$}
  \psfrag{-1}{$-1.0$}
  \psfrag{ 1}{1}
  \psfrag{ 10}{10}
  \psfrag{ 100}{$10^2$}
  \psfrag{ 1000}{$10^3$}
  \psfrag{ 10000}{$10^4$}
  \psfrag{ 100000}{$10^5$}
  \psfrag{ 1000000}{$10^6$}
  \includegraphics[scale=0.8]{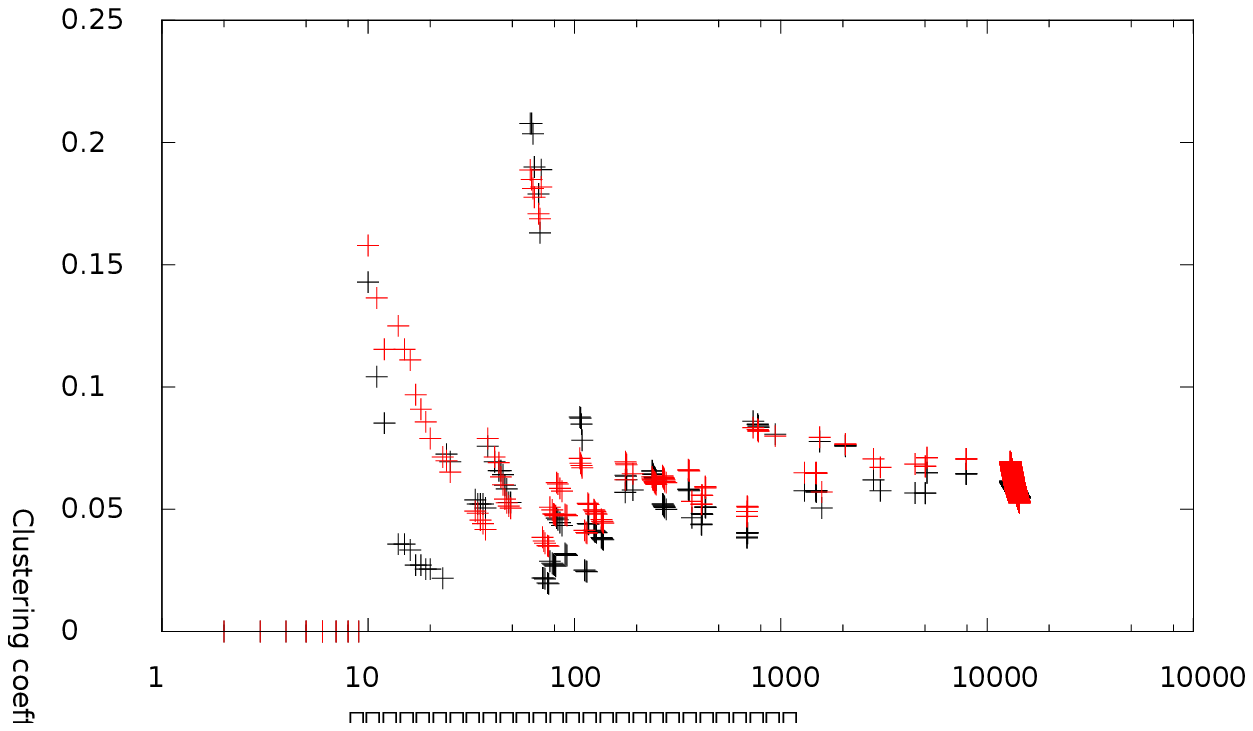}
  \includegraphics[scale=0.8]{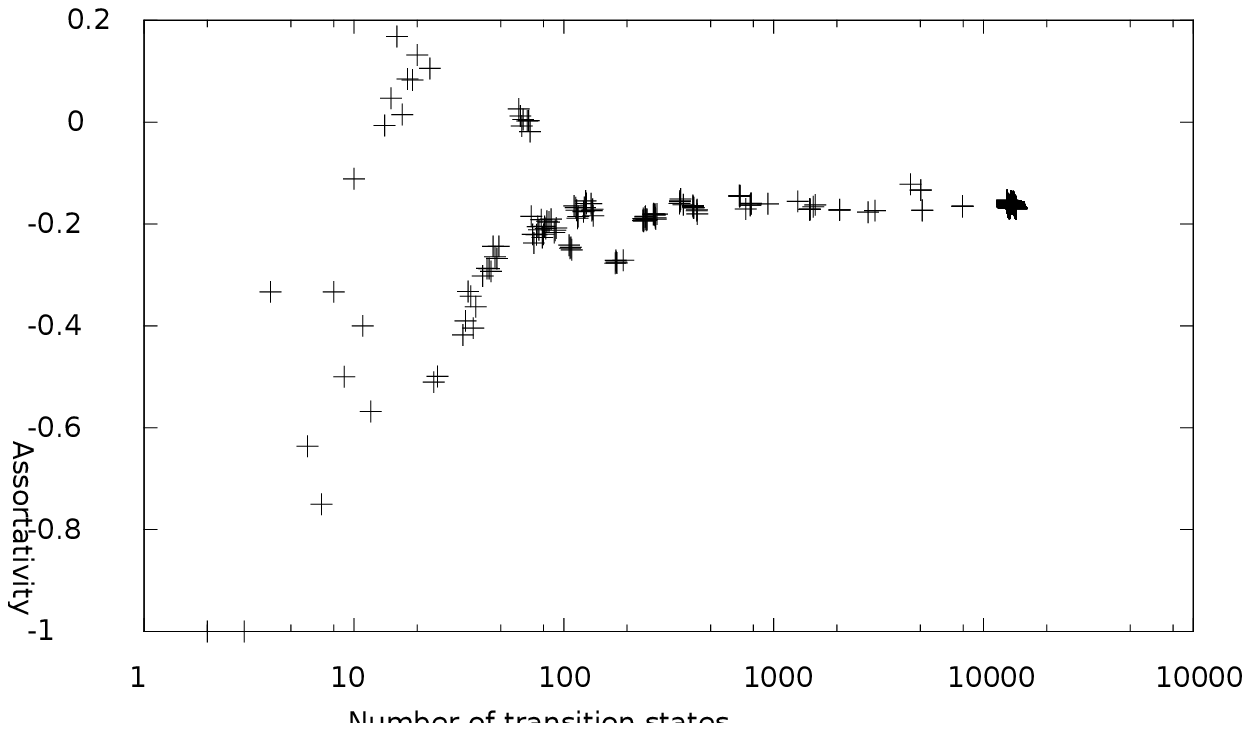}
  \caption{Top: variation of the average clustering coefficient (black) and
transitivity (red) as the largest connected component of the BLJ database
grows. Bottom: the corresponding plot for the assortativity.}
  \label{fig:convergenceBLJ}
\end{figure}

\begin{figure}[ht]
  \centering
  \psfrag{Number of transition states}{Number of transition states}
  \psfrag{Ratio}{Ratio}
  \psfrag{ 0.1}{$10^{-1}$}
  \psfrag{ 1}{1}
  \psfrag{ 10}{10}
  \psfrag{ 100}{$10^2$}
  \psfrag{ 1000}{$10^3$}
  \psfrag{ 10000}{$10^4$}
  \psfrag{ 100000}{$10^5$}
  \includegraphics[scale=0.8]{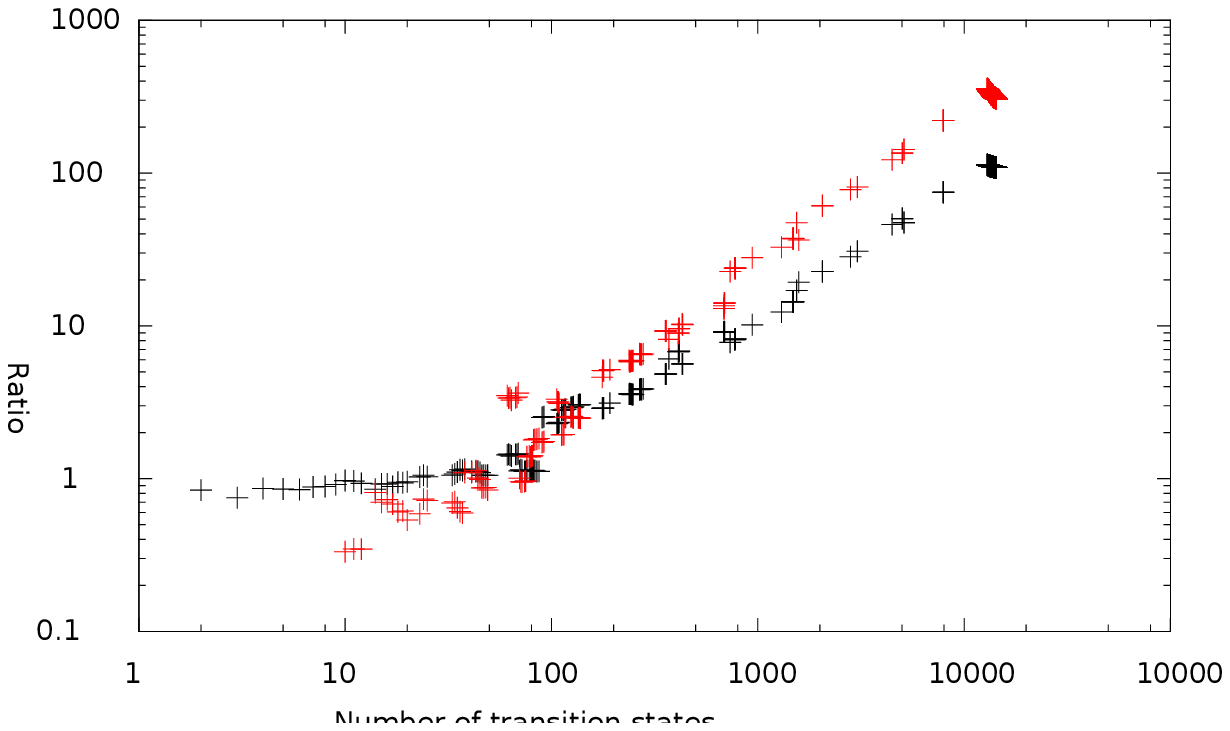}
  \caption{The ratio of average shortest path length for the largest connected
component of the BLJ database to that of the equivalent random graph (black)
and the corresponding ratio for the transitivity (red).}
  \label{fig:smallWorldBLJ}
\end{figure}

Our analysis followed the steps described above for OTP. The convergence of the average
clustering coefficient, transitivity and assortativity are shown in figure
\ref{fig:convergenceBLJ}, and the ratios of the average shortest path length
and transitivity to those for the equivalent random graph are shown in figure
\ref{fig:smallWorldBLJ}. The results are qualitatively similar to OTP, although
the convergence of the average clustering coefficient and transitivity is less
clear due to the smaller sample size. The value of the assortativity is
negative, indicating slight disassortativity compared to the slight
assortativity of the OTP system. Park and Newman suggest that disassortativity
can arise due to the exclusion of alternative and self-connections,\cite{ParkN03}
so this result is unsurprising, although it is unclear why the opposite
behaviour to OTP is observed. The growing ratios of average shortest path
length and transitivity to the equivalent random graph suggest the same
conclusion: that the system is showing lattice graph behaviour rather than
small-world behaviour.

\section{Conclusions}
\label{sec:conc}
Networks of minima and transition states for Morse clusters with different
ranges in the interatomic potential were analysed. The networks exhibited
small-world properties, but statistical testing using recent methods
demonstrated that they are probably not scale-free. The $\mathrm{LJ}_{14}$
cluster was re-evaluated using these new statistical methods and the tests were
inconclusive: we cannot firmly conclude the network is scale-free, but it
remains a possibility.

The Morse networks were compared with glassy systems,
which have very different energy landscapes characterised by numerous
low energy amorphous structures separated by relatively high barriers.
The two glassy networks did not show small-world behaviour, which suggests that the small-world properties of
the LJ and Morse networks are a feature of their single-funnel
potential energy landscapes.
However, further investigation is required. Larger clusters with more minima
and transition states will make statistical tests for scale-free behaviour more
reliable.

Using the convergence of network properties to test whether a sample
is representative opens the possibility of studying systems where it is
impractical to find the complete network.
While it remains computationally intractable to find the complete
network for glassy systems, as the number of minima and transition states is
far too large, we believe the approach here will allow further insight for
such systems.
It is certainly possible to generate and analyse larger
samples, and if databases generated by different schemes
converge on the same properties we can have greater confidence about whether the
samples are truly representative. Analysing the network properties of a
representative sample will provide deeper understanding of the behaviour of bulk
glasses.
The $\mathrm{LJ}_{38}$ cluster, which
has a double-funnel landscape,\cite{DoyeMW99a} could produce interesting
results if the two funnels are considered separately and together. Studying a
fast-folding protein network would also be insightful. On the basis of the
results here, we anticipate that the folding network is not really scale-free,
but due to its known structure-seeking nature, small-world character is likely.

\bigskip

\leftline{\bf Conflicts of Interest}
There are no conflicts of interest to declare.

\leftline{\bf Acknowledgements}
JM acknowledges the support of a Sackler Studentship from the University of Cambridge.
Part of this work was performed while DM was a member of the Department of Chemistry at the University of Cambridge,
financially supported by the Engineering and Physical Sciences
Research Council and the European Research Council. Data can be accessed at
\texttt{http://doi.org/10.5281/zenodo.580085}.

\def\acr{Acc. Chem. Res.}
\def\acrys{Acta. Crystallogr.}
\def\aca{Acta. Crystallogr. A}
\def\apsu{Acta. Physicochim. URS.}
\def\acp{Adv. Chem. Phys.}
\def\am{Adv. Mater.}
\def\ap{Adv. Phys.}
\def\apc{Adv. Prot. Chem.}
\def\acsn{ACS Nano}
\def\ajp{Am. J. Phys.}
\def\aciee{Angew. Chem. Int. Edit.}
\def\ap{Ann. Physik}
\def\arpc{Ann. Rev. Phys. Chem.}
\def\apl{Appl. Phys. Lett.}
\def\bjp{Braz. J. Phys.}
\def\cp{Chem. Phys.}
\def\cpl{Chem. Phys. Lett.}
\def\crev{Chem. Rev.}
\def\cccc{Coll. Czech. Chem. Comm.}
\def\cosu{Colloid Surface}
\def\cpc{Comp. Phys. Comm.}
\def\cphyc{ChemPhysChem}
\def\cj{Comput. J.}
\def\comp{Computing}
\def\cocis{Curr. Opin. Colloid In.}
\def\cossms{Curr. Opin. Solid St. M.}
\def\csab{Calcutta Statist. Assoc. Bull.}
\def\econ{Econometrica}
\def\el{Europhys. Lett.}
\def\fd{Faraday Disc.}
\def\ic{Inorg. Chem.}
\def\ijmpc{Int. J. Mod. Phys. C}
\def\ijqc{Int. J. Quant. Chem.}
\def\irpc{Int. Rev. Phys. Chem.}
\def\jacs{J. Am. Chem. Soc.}
\def\jap{J. Appl. Phys.}
\def\jas{J. Atmos. Sci.}
\def\jbc{J. Biol. Chem.}
\def\jce{J. Chem. Educ.}
\def\jcp{J. Chem. Phys.}
\def\jcscc{J. Chem. Soc., Chem. Commun.}
\def\dalton{J. Chem. Soc., Dalton Trans.}
\def\fdisc{J. Chem. Soc., Faraday Discuss.}
\def\faraday{J. Chem. Soc. Faraday T.}
\def\faradi{J. Chem. Soc. Farad. T. I}
\def\jcis{J. Colloid Interface Sci.}
\def\jcc{J. Comp. Chem.}
\def\jcop{J. Comp. Phys.}
\def\jcr{J. Confl. Resolut.}
\def\jcry{J. Cryst. Growth}
\def\jetp{J. Exp. Theor. Phys. (Russia)}
\def\jgo{J. Global Optim.}
\def\jlcm{J. Less-Common Met.}
\def\jmch{J. Mater. Chem.}
\def\jmc{J. Math. Chem.}
\def\jmb{J. Mol. Biol.}
\def\jmm{J. Mol. Model.}
\def\jmsp{J. Mol. Spec.}
\def\jmst{J. Mol. Struct.}
\def\jmstt{J. Mol. Struc-Theochem}
\def\jncs{J. Non-Cryst. Solids}
\def\jpc{J. Phys. Chem.}
\def\jpca{J. Phys. Chem. A}
\def\jpcb{J. Phys. Chem. B}
\def\jpa{J. Phys. A}
\def\jpb{J. Phys. B}
\def\jpcssp{J. Phys. C: Solid State Phys.}
\def\jpcs{J. Phys. Chem. Solids}
\def\jpcm{J. Phys. Condens. Mat.}
\def\jpfmp{J. Phys. F, Metal Phys.}
\def\jpsj{J. Phys. Soc. Jpn.}
\def\jrnist{J. Res. Natl. Inst. Stand. Technol.}
\def\jsp{J. Stat. Phys.}
\def\lang{Langmuir}
\def\ledpm {Lond. Edinb. Dubl. Phil. Mag.}
\def\mcp{Macromol. Chem. Physic.}
\def\mrsb{Mater. Res. Soc. Bull.}
\def\msr{Mater. Sci. Rep.}
\def\mcom{Math. Comput.}
\def\mg{Math. Gazette}
\def\mp{Mol. Phys.}
\def\nl{Nano Lett.}
\def\nanos{Nanoscale}
\def\nat{Nature}
\def\nm{Nat. Mater.}
\def\nsb{Nat. Struct. Biol.}
\def\nrlq{Nav. Res. Logist. Q.}
\def\njc{New J. Chem.}
\def\numa{Numer. Math.}
\def\phm{Philos. Mag.}
\def\pma{Philos. Mag. A}
\def\pml{Philos. Mag. Lett.}
\def\ptrs{Philos. T. Roy. Soc.}
\def\ptrsa{Philos. T. Roy. Soc. A}
\def\ptrsb{Philos. T. Roy. Soc. B}
\def\pccp{Phys. Chem. Chem. Phys.}
\def\pla{Phys. Lett. A}
\def\plb{Phys. Lett. B}
\def\prep{Phys. Reports}
\def\pr{Phys. Rev.}
\def\pra{Phys. Rev. A}
\def\prb{Phys. Rev. B}
\def\prbcm{Phys. Rev. B}
\def\prc{Phys. Rev. C}
\def\prd{Phys. Rev. D}
\def\pre{Phys. Rev. E}
\def\prl{Phys. Rev. Lett.}
\def\pss{Phys. State Solidi}
\def\pssb{Phys. State Solidi B}
\def\Pa{Physica A}
\def\phys{Physics}
\def\pnas{Proc. Natl. Acad. Sci. USA}
\def\pnasu{Proc. Natl. Acad. Sci. USA}
\def\prs{Proc. R. Soc.}
\def\prsa{Proc. R. Soc. A}
\def\prslam{P. Roy. Soc. Lond. A Mat.}
\def\psfg{Proteins: Struct., Func. and Gen.}
\def\psych{Psychol. Today}
\def\psycho{Psychometrika}
\def\pmd{Publ. Math. Debrecen}
\def\pac{Pure. Appl. Chem.}
\def\rpp{Rep. Prog. Phys.}
\def\rmp{Rev. Mod. Phys.}
\def\sci{Science}
\def\siamr{SIAM Review}
\def\smat{Soft Matter}
\def\spj{Sov. Phys. JETP}
\def\ss{Surf. Sci.}
\def\tetra{Tetrahedron}
\def\tca{Theor. Chim. Acta}
\def\tfs{Trans. Faraday Soc.}
\def\zp{Z. Phys.}
\def\zpb{Z. Phys. B.}
\def\zpc{Z. Phys. Chem.}
\def\zpcb{Z. Phys. Chem. Abt. B}
\def\zpd{Z. Phys. D}
\def\zfpd{Z. Phys. D}
\def\zpdamc{Z. Phys. D}

%\nolinenumbers

%This is where your bibliography is generated. Make sure that your .bib file is actually called library.bib
\bibliography{networks}

%This defines the bibliographies style. Search online for a list of available styles.
\bibliographystyle{ieeetr}

\end{document}